# Evidence for time-reversal symmetry breaking and glassy dynamics in superconducting $(La,Pr,Sm)_3Ni_2O_7$ films

Haoran Ji[1,2#], Zheyuan Xie[1,2#], Yaqi Chen[3#], Guangdi Zhou[4,5#], Keyi Liang[6], Rui Liu[6], Mengqiao Geng[1], Meixuan Li[1], Longxin Pan[7,8], Heng Wang[4,5], Haoliang Huang[4,5], Jun Ge[1,2], Yi Liu[7,8], Yang Liu[1,2], Guang-Ming Zhang[9,10], Yihua Wang[6], Ziqiang Wang[11*], Qi-Kun Xue[4,5,10], Zhuoyu Chen[4,5*] & Jian Wang[1,2,12,13*]

[1]International Center for Quantum Materials, School of Physics, Peking University, Beijing 100871, China

[2]Beijing Key Laboratory of Quantum Devices, Peking University, Beijing 100871, China

[3]College of Semiconductors (National Graduate College for Engineers), Southern University of Science and Technology, Shenzhen 518055, China

[4]State Key Laboratory of Quantum Functional Materials, Department of Physics, and Guangdong Basic Research Center of Excellence for Quantum Science, Southern University of Science and Technology, Shenzhen 518055, China

[5]Quantum Science Center of Guangdong-Hong Kong-Macao Greater Bay Area, Shenzhen, 518045, China

[6]State Key Laboratory of Surface Physics and Department of Physics, Fudan University, Shanghai, China.

[7]School of Physics and Beijing Key Laboratory of Opto-electronic Functional Materials & Micro-nano Devices, Renmin University of China, Beijing 100872, China

[8]Key Laboratory of Quantum State Construction and Manipulation (Ministry of Education), Renmin University of China, Beijing 100872, China

[9]School of Physical Science and Technology, ShanghaiTech University, Shanghai 201210, China

[10]Department of Physics, Tsinghua University, Beijing 100084, China

[11]Department of Physics, Boston College, Chestnut Hill, MA 02467, USA

[12]Collaborative Innovation Center of Quantum Matter, Beijing 100871, China

[13]Hefei National Laboratory, Hefei 230088, China

[#]These authors contribute equally.

*Correspondence to: jianwangphysics@pku.edu.cn (J.W.),
chenzhuoyu@sustech.edu.cn (Z.C.),
wangzi@bc.edu (Z.W.)

**Superconductivity and magnetism are generally distinct and competing orders, characterized by different broken symmetries and order parameters. Although they may coexist in some correlated systems, whether and how these orders can become inherently intertwined in unconventional high-transition-temperature (high-$T_c$) superconductors remains a central open question. Here, we report time-reversal symmetry (TRS) breaking and glassy dynamics in the superconducting state of nickelate (La,Pr,Sm)$_3$Ni$_2$O$_7$ films. Magnetoresistance hysteresis and zero-field non-reciprocal charge transport, complementary evidence for broken TRS, are observed in the superconducting state, while absent in the normal state. Within the superconducting transition regime, logarithmically slow resistance relaxations following magnetic-field perturbations provide strong evidence for glassy dynamics. Remarkably, the reproducible temperature hierarchy across multiple films reveals that these novel phenomena emerge concurrently with developing superconducting phase coherence, preceding the zero-resistance state. Consistently, scanning superconducting quantum interference device (sSQUID) microscopy directly resolves the zero-field magnetic-flux patterns indicative of spontaneous TRS breaking, which emerge simultaneously with local superconducting diamagnetism and phase stiffness. Our observations suggest a frustrated multicomponent superconducting order parameter in the nickelate films, whose internal relative phases may carry TRS-breaking chirality and freeze into a textured phase glass. Collectively, our findings identify a distinctive form of superconductivity in which internal phase degrees of freedom may support additional broken symmetries and glassy order, offering a new perspective on intertwined ordering phenomena in unconventional high-$T_c$ superconductors.**

Since the discovery of high-transition-temperature (high-$T_c$) superconductivity in the cuprates[1,2], layered oxides that are structural and electronic analogs of the cuprates have been pursued for decades as potential candidates for high-$T_c$ superconductivity[3]. Extending this strategy, superconductivity with $T_c$ around 15 K was achieved in the infinite-layer nickelate $R_{1-x}$Sr$_x$NiO$_2$ ($R$ = La, Pr, Nd, …) thin films[4], which share the $3d^9$ electronic configuration (Ni$^+$) and are isostructural to the cuprates. Subsequently, in the Ruddlesden-Popper (R-P) nickelates $R_{n+1}$Ni$_n$O$_{3n+1}$ ($n$ = 1, 2, 3,…,∞), superconductivity with onset $T_c$ near 80 K was discovered in bilayer (La, Pr)$_3$Ni$_2$O$_7$ and near 30 K in trilayer La$_4$Ni$_3$O$_{10}$ bulk crystals under high pressure[5-9], establishing nickelates as a new family of high-$T_c$ superconductors[10,11]. Structurally, the R-P bilayer nickelates host covalently bonded apical oxygen, which couples the adjacent Ni-O layers, and stabilizes a mixed-valence state of Ni$^{2.5+}$ with $3d^{7.5}$ configuration. The bilayer nickelates involve both $3d_{z^2}$ and $3d_{x^2-y^2}$ orbitals near the Fermi level[5,12-19] and necessitate the consideration of interlayer coupling[20-23], differing from the single-orbital picture of the cuprates. The bilayer nickelate superconductors with distinct electronic structures and strong electronic correlations immediately sparked extensive research interest. However, the requirement of high pressure has made it very challenging to study the properties of the superconducting (SC) state using techniques such as angle-resolved photoemission spectroscopy (ARPES)[24-26], scanning tunneling microscopy/spectroscopy (STM/STS)[27], and scanning superconducting quantum interference device (sSQUID) microscopy. Recently, superconductivity with onset $T_c$ above 40 K was achieved in (La, Pr)$_3$Ni$_2$O$_7$ thin films at ambient pressure via epitaxial compressive strain[28-31]. This thin-film platform enables direct characterization through versatile experimental techniques, thereby providing an exceptional opportunity to investigate the nature of the SC state.

Here, through systematic electrical transport and sSQUID measurements, we report a distinctive SC state with spontaneous time-reversal symmetry (TRS) breaking and glassy dynamics in the R-P bilayer nickelate (La, Pr, Sm)$_3$Ni$_2$O$_7$ films. With decreasing temperature, these TRS-breaking and glassy signatures emerge spontaneously just in the lower-temperature portion of SC transition regime below a characteristic temperature $T^*$, with no detectable signatures in its higher-temperature portion or in the normal state. First, magnetoresistance hysteresis emerges in this lower-temperature SC transition regime and persists into the zero-resistance state, suggesting TRS breaking within the SC state. The hysteresis and zero-resistance superconductivity are weakened simultaneously upon progressive oxygen reduction, suggesting a cooperative relation rather than competition between them. Second, non-reciprocal charge transport, manifested as asymmetric current-voltage responses for opposite current directions, is observed at zero magnetic field in the SC state, providing complementary evidence for TRS breaking at zero field. Third, logarithmic time-dependent resistance relaxations are observed after external magnetic-field perturbations, revealing the glassy dynamics. The onset temperature of glassy dynamics tends to increase with improved sample quality and enhanced superconductivity, suggesting that the glassiness is not due to extrinsic disorder. Fourth, sSQUID magnetometry resolves zero-field magnetic-flux patterns in the SC state, providing direct evidence for spontaneous TRS breaking. The magnetic flux and the local SC diamagnetism exhibit nearly coincident onset temperatures, linking the emergence of TRS breaking to the development of local SC phase stiffness. Consistent with this correspondence, transport measurements across multiple films establish a common temperature hierarchy in which TRS breaking and glassy behavior emerge in the lower-temperature portion of SC transition regime, where finite-scale SC phase coherence is developing below $T^*$. Collectively, these observations demonstrate that spontaneous TRS breaking and glassiness are intrinsically intertwined with superconductivity in bilayer nickelate (La, Pr, Sm)$_3$Ni$_2$O$_7$ films, and point to a common origin in the SC phase sector. One possible microscopic scenario is a frustrated multicomponent SC order parameter, in which internal relative-phase textures develop together with the overall SC phase stiffness and give rise to the observed TRS breaking and glassy behavior.

## Two-dimensional superconductivity

The R-P bilayer nickelate (La, Pr, Sm)$_3$Ni$_2$O$_7$ has a stacking structure of (La, Pr, Sm)O-NiO$_2$-(La, Pr, Sm)O-NiO$_2$-(La, Pr, Sm)O, as schematically shown in Fig. 1**a**. The pure-phase single-crystal (La, Pr, Sm)$_3$Ni$_2$O$_7$ films are grown on the (001)-oriented SrLaAlO$_4$ substrates by gigantic-oxidative atomic layer-by-layer epitaxy[29,32]. To mitigate the interfacial structural discontinuity, an insulating (La, Pr)O-NiO$_2$-(La, Pr)O buffer layer is incorporated between the film and substrate. The crystalline quality of a three-unit-cell (3-UC) thick La$_{1.95}$Pr$_{1.05}$Ni$_2$O$_7$ thin film is demonstrated by the atomic-resolution scanning transmission electron microscopy (STEM) images (Fig. 1**b** and **c**), where no impurity phases or discernible cation impurities are detected. The in-plane lattice constant of bulk La$_{2.85}$Pr$_{0.15}$Ni$_2$O$_7$ is ~ 3.832 Å, so that the thin films grown on (001)-oriented SrLaAlO$_4$ substrate (3.75 Å) experience a roughly 2% compressive strain. It is suggested that the short in-plane lattice constant (i.e., short in-plane Ni-O bond length) is crucial for the ambient-pressure superconductivity in the bilayer nickelates[28,33].

The electrical transport measurements are performed on a 3-UC La$_{1.95}$Pr$_{1.05}$Ni$_2$O$_7$ thin film on SrLaAlO$_4$ substrate with a buffer layer (sample S1), using standard electrodes with Hall-bar geometry (right inset of Fig. 1**d**). The temperature-dependent sheet resistance $R_S(T)$ curve shows a smooth superconducting transition (Fig. 1**d**), with an onset around $T_c^{onset}$ = 42.1 K and a zero-resistance state below $T_c^{zero}$ = 24.9 K. The $R_S(T)$ curve deviates from its normal-state extrapolation around 48.4 K, defining the normal-state sheet resistance $R_N$. The smooth tail of the $R_S(T)$ curve approaching zero resistance is captured by the Berezinskii-Kosterlitz-Thouless (BKT) transition in two dimensions[34,35]. Fitting this $R_S(T)$ tail with the Halperin-Nelson equation[36] yields a transition temperature $T_{BKT}$ = 24.5 K (left inset of Fig. 1**d**). Complementary evidence for the BKT transition is provided by the current-voltage ($I$-$V$) curves, which exhibit a power-law dependence $V \propto I^{\alpha}$ (inset of Fig. 1**e**). The extracted exponent $\alpha$ increases rapidly with decreasing temperature, and reaches 3 at 22.3 K, yielding a comparable $T_{BKT}$ (Fig. 1**e**).

Upon cooling through the BKT transition, free vortices–topological phase singularities of the complex SC order parameter–bind into vortex-antivortex pairs, establishing quasi-long-range phase coherence and the zero-resistance SC state. To characterize the preceding development of phase coherence within the SC transition regime, we introduce a characteristic temperature, $T^* = 35$ K, operationally defined as the temperature below which the *I*-*V* characteristics first become detectably nonlinear ($V \propto I^{\alpha}$, with $\alpha$ exceeding unity, as shown in Fig. 1**e**). Within the BKT framework, such nonlinearity arises from the current-driven unbinding of vortex-antivortex pairs, and the exponent is related to the renormalized SC phase stiffness $J_s$ through[37,38]

$$\alpha(T) = 1 + \frac{\pi J_s(T)}{T}.$$

The emergence of nonlinear *I*-*V* behavior therefore provides an experimental signature of developing SC phase rigidity on a finite length scale[38,39]. Located between $T_c^{onset}$ and $T_c^{zero}$, $T^*$ thus marks a crossover into a regime of appreciable finite-scale SC phase coherence and phase rigidity, preceding the establishment of global phase coherence. As shown below, the TRS-breaking and glassy phenomena are observed predominantly below $T^*$.

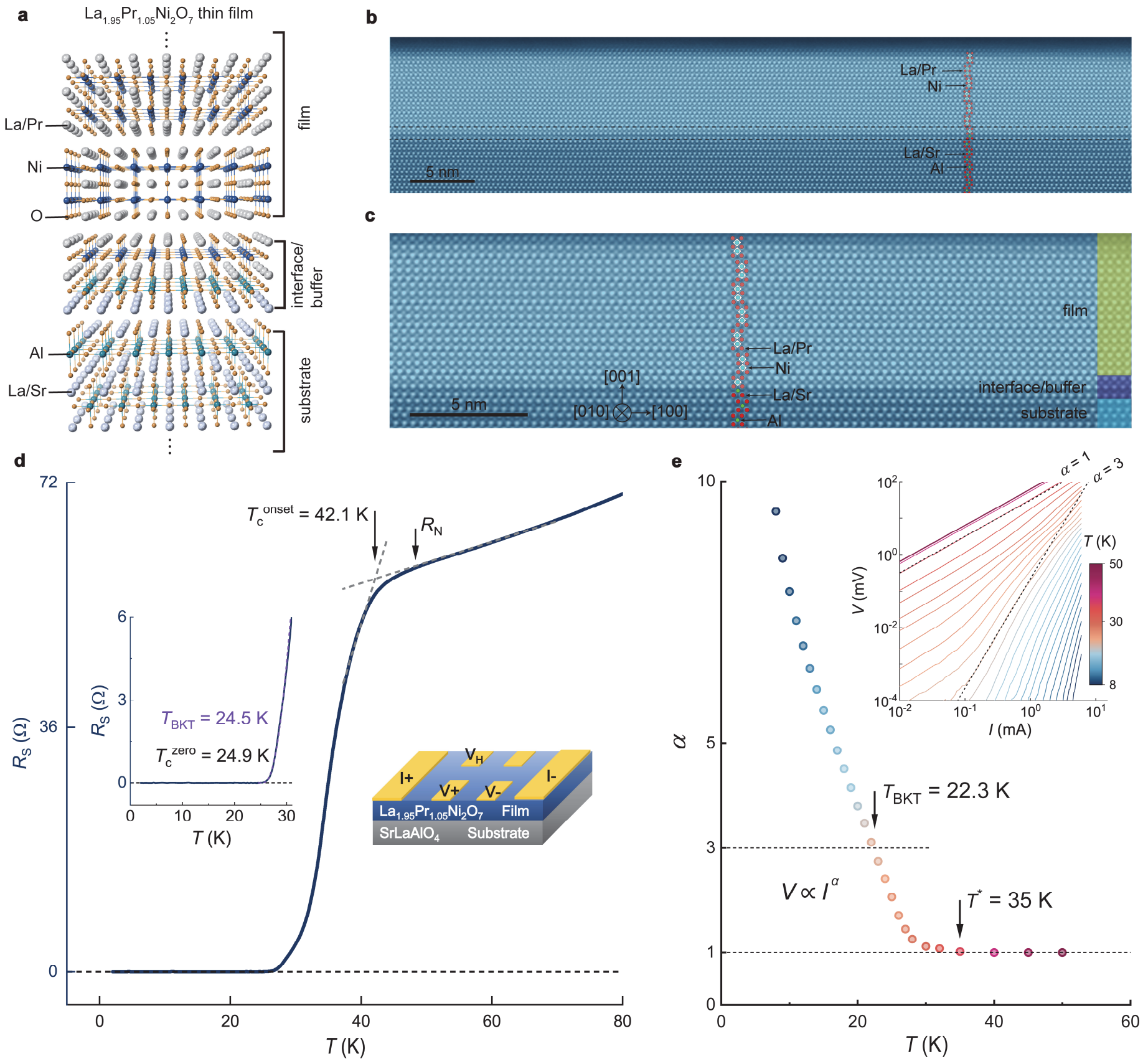

**Fig. 1 | Two-dimensional superconductivity in bilayer nickelate $La_{1.95}Pr_{1.05}Ni_2O_7$ thin film (sample S1). a**, Crystal structure of the bilayer nickelate $La_{1.95}Pr_{1.05}Ni_2O_7$ film grown on (001)-$SrLaAlO_4$ substrate with a buffer layer. **b,** Scanning transmission electron microscopy (STEM) image of a $La_{1.95}Pr_{1.05}Ni_2O_7$ film with a large field of view. Two black dashed lines indicate the interface/buffer layer. **c,** Zoomed-in view of the STEM image from **b**. **d**, Temperature-dependent sheet resistance $R_S(T)$ curve of the

$La_{1.95}Pr_{1.05}Ni_2O_7$ thin film (S1). $T_c^{onset} = 42.1$ K is defined by the crossing point between the extrapolated normal state and SC transition. $R_N$ is defined at the deviation from the normal-state extrapolation. Left inset exhibits the enlarged view of $R_S(T)$ curve, showing a zero-resistance transition at $T_c^{zero} = 24.9$ K. Purple dashed line represents the fit curve for Berezinskii-Kosterlitz-Thouless (BKT) transition, yielding a BKT temperature $T_{BKT} = 24.5$ K. Right inset shows the measurement configurations with Hall bar geometry. **e,** Temperature-dependence of $\alpha$, determined by $V \propto I^{\alpha}$ from the inset. With increasing temperatures, $\alpha$ drops to 3 at $T_{BKT} = 22.3$ K. The characteristic temperature $T^* = 35$ K is defined by $\alpha$ just exceeding 1. Inset shows $I$-$V$ curves from 8 K to 50 K. Two black dashed lines mark $V \propto I^{\alpha = 3}$ and $V \propto I^{\alpha = 1}$, respectively.

**Hysteresis in magnetoresistance under out-of-plane magnetic fields**

With out-of-plane magnetic field $B_{\perp}$ sweeping along opposite directions, evident hysteretic behaviors are observed in the magnetoresistance $R_S(B)$ curves in the lower-temperature portion of the SC transition regime below $T^*$, persisting to temperatures below $T_c^{zero}$ (Fig. 2**a**). Figures 2**d-e** show the representative $R_S(B)$ curves at 24 K at two scales. The two branches of the $R_S(B)$ curves, measured with magnetic fields sweeping from 14 T to −14 T (blue dashed line) and −14 T to 14 T (red solid line), overlap for magnetic-field magnitudes larger than approximately 2 T. However, the two branches diverge and manifest a distinct hysteresis loop around 0 T (Fig. 2**d** and **e**). Furthermore, each branch of $R_S(B)$ curve is asymmetric with respect to $B_{\perp} = 0$ T. Such a history-dependent magnetoresistance behavior, irreversible under the time-reversal operation, suggests TRS breaking in the SC state. With increasing temperatures, the hysteresis weakens, and becomes indiscernible above a characteristic temperature $T_{hys}$ of 30 K (Upper panels in Fig. 2**a**, $T_{hys} < T^*$). Representative $R_S(B)$ curves at 36 K are shown in Fig. 2**b-c**, where the two branches show no discernible difference, even when zooming-in down to 0.1 T, confirming the absence of hysteresis in the higher-temperature regime of SC transition. It should be emphasized that the magnetic fields are calibrated by a high-resolution InAs Hall sensor (denoted as $B'$ in the figures), which precludes the extrinsic issues such as the remnant field of the measurement system. Notably, the hysteresis loop exhibits coalescing minima, where the minima of two $R_S(B)$ traces converge at $B = 0$ T, in sharp contrast to the two split minima typically associated with coercive field in long-range-ordered ferromagnetic systems[40-42]. The hysteresis also persists across different magnetic-field orientations, even for in-plane fields (Extended Data Figs. 3 and 4), which argues against a trapped-vortex scenario.

To quantify the hysteresis, we define the horizontal displacement $\Delta B$ between the two branches of $R_S(B)$ curves: $\Delta B(R_S) = B^{\downarrow}(R_S) - B^{\uparrow}(R_S)$, where ↓ denotes field sweep from 14 T to −14 T, and ↑ in reverse (schematically marked by horizontal arrows in Fig. 2**e**).

The maximum magnitude, $\Delta B_{\max}$, provides a measure for hysteresis: it emerges at $T_{\mathrm{hys}}$ of 30 K, increases monotonically as temperature decreases, and persists below $T_c^{\mathrm{zero}}$ (Fig. 2**f**). The temperature evolution of hysteresis demonstrates that the TRS-breaking signatures emerge specifically in the lower-temperature portion of the SC transition below $T^{*}$, and extend to temperatures below $T_c^{\mathrm{zero}}$.

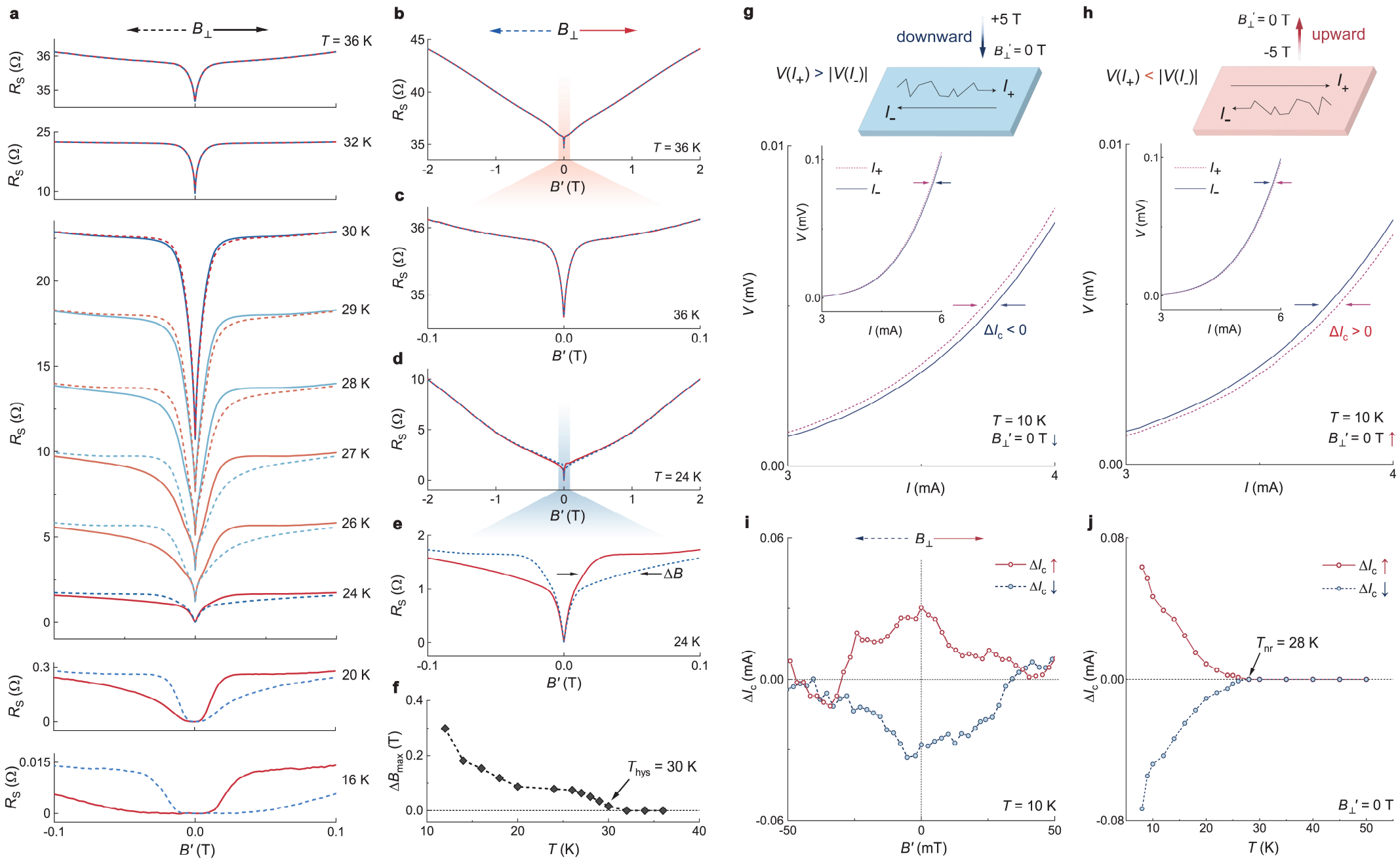

**Fig. 2 | Magnetoresistance hysteresis and non-reciprocal charge transport in $La_{1.95}Pr_{1.05}Ni_2O_7$ thin film (S1). a**, Magnetoresistance $R_S(B)$ curves at different temperatures, shown within $|B_{\perp}| \leq 0.1$ T. Curves are offset vertically for clarity. Here, all $R_S(B)$ curves are measured with out-of-plane field $B_{\perp}$ sweeping from 14 T to −14 T (dashed lines) and −14 T to 14 T (solid lines), while the selected range of $R_S(B)$ curves is displayed for clarity. A high-resolution Hall sensor is used to calibrate the magnetic fields within 5 T, and the calibrated magnetic field is denoted as $B'$. **b** and **c**, Representative $R_S(B)$ curves from **a** at $T = 36$ K with $|B_{\perp}|$ ranging within 2 T (**b**) and 0.1 T (**c**), showing no discernible hysteresis. **d** and **e**, Representative $R_S(B)$ curves from **a** at $T = 24$ K with $|B_{\perp}|$ ranging within 2 T (**d**), 0.1 T (**e**), showing evident hysteresis loop. Black arrows schematically mark $\Delta B$. The shadows denote the zoomed-in ranges. **f,** Temperature-dependence of hysteresis width $\Delta B_{\max}$. The hysteresis emerges below $T_{\mathrm{hys}} = 30$ K. **g** and **h,** *I*-*V* curves at 10 K at zero field, containing both $V(I_{+})$ (Purple dashed curve, current along positive direction) and $|V(I_{-})|$ (Dark blue solid curve, current along negative direction) branches. Opposing arrows highlight the deviation between $V(I_{+})$ and $|V(I_{-})|$ curves, i.e., non-reciprocal response, while the color arrangement of the arrows reflects the non-reciprocal polarity. Insets in **g** and **h** show the *I*-*V* curves over a wider current range. Schematics above **g** and **h** illustrate the magnetic field history-dependent non-reciprocal charge transport at zero external magnetic field. Left: Following a downward field-swept history (from +5 T to 0 T), the

voltage magnitude is larger for positive current direction than for negative direction, i.e., $V(I_+) > |V(I_-)|$. The SC film is more resistive under $I_+$ (polyline) than under $I_-$ (straight line). Right: Following an upward field-swept history (from - 5 T to 0 T), the non-reciprocal polarity reverses, $V(I_+) < |V(I_-)|$, as schematically indicated by the exchange of the two line shapes. **i,** Non-reciprocal critical current $\Delta I_c$ as a function of $B'$. $\Delta I_c = I_{c+} - |I_{c-}|$, defined at $V$ = 0.005 mV. **j,** Temperature-dependence of zero-field $\Delta I_c$ at two polarities. The non-reciprocal transport emerges below $T_{nr}$ = 28 K. Arrows appended to $B'$ and $\Delta I_c$ denote the field history (red ↑: upward sweep; blue ↓: downward sweep).

**Non-reciprocal transport characteristics**

In parallel to magnetoresistance hysteresis, non-reciprocal charge transport is observed in the SC state in $La_{1.95}Pr_{1.05}Ni_2O_7$ film at zero external magnetic field, providing complementary evidence for spontaneous symmetry breakings[43-48]. Figures 2**g** and **h** show the $V(I_+)$ and $|V(I_-)|$ curves at 10 K under zero field, where $I_+(I_-)$ denotes the current applied in the positive (negative) direction, respectively. In the finite-resistance regime, $V(I_+)$ and $|V(I_-)|$ curves exhibit evident splits: the voltage amplitudes differ, $V(I_+) \neq |V(I_-)|$, when currents of equal magnitude ($I_+ = |I_-|$) are applied along opposite directions. This directional asymmetry in electrical properties is referred to as non-reciprocal charge transport, which at zero external field requires spontaneous TRS breaking in a SC system[43-48]. Here, the zero-field condition is achieved by two different magnetic field-histories: either downward sweeping from 5 T to 0 T (Fig. 2**g**) or upward sweeping from -5 T to 0 T (Fig. 2**h**). The magnetic field is applied along the out-of-plane direction, and its strength ($B'$) is calibrated by the Hall sensor. It should be noted that the retraceable $V(I_+)$ overlaps nicely with $V(I_+$ P-0$)$ (current scans from positive to zero) and $|V(I_-)|$ overlaps with $|V(I_-$ N-0$)|$ (current scans from negative to zero), which eliminates the heating-induced artifact (Supplementary Fig. S10). Furthermore, the $V(I_+)$ and $|V(I_-)|$ curves overlap with each other, i.e., the non-reciprocity is absent, at high temperatures above 30 K or under large magnetic fields above 5 T (Supplementary Fig. S11). These controlled results confirm the validity of our measurements and support the intrinsic nature of non-reciprocal characteristics in the SC regime below $T^*$.

The non-reciprocal *I-V* characteristics depend strongly on the magnetic field history, where $B_\perp$ is swept along two opposite directions: upward sweep from -5 T to +5 T and downward sweep in reverse. For convenience, we define the non-reciprocal critical current $\Delta I_c = I_{c+} - |I_{c-}|$, with $I_{c+}$ and $I_{c-}$ corresponding to SC critical currents at which the measured voltages reach 0.005 mV, a small value slightly above the noise floor of zero resistance. The obtained magnetic field-dependence of non-reciprocal critical current $\Delta I_c(B)$ exhibits a prominent hysteresis loop over a field range of approximately 50 mT (Fig. 2**i**). The hysteresis loop in $\Delta I_c(B)$ ensures that there are two separate and, most importantly, non-zero $\Delta I_c$ values at 0 T, supporting the spontaneous TRS breaking in the absence of external magnetic field. Hysteretic $\Delta I_c(B)$ behaviors were also observed in SC diode effect based on superconductor/ferromagnet multilayers[49].

The polarity of non-reciprocal signals, defined by the positive or negative sign of $\Delta I_c$, is reversible and controllable by magnetic field training. Specifically, the polarity is negative after a downward field sweep to zero field, and positive after an upward sweep. The reversibility argues against extrinsic origins such as the fixed thermal gradient or thermoelectric effect induced by electrodes. For both polarities, the magnitudes of zero-field $\Delta I_c$ decrease with increasing temperature and vanish around $T_{nr}$ = 28 K (Fig. 2**j**), close to the temperature where magnetoresistance hysteresis disappears. Collectively, the mutually consistent non-reciprocal *I-V* characteristics and magnetoresistance hysteresis provide complementary evidence for TRS breaking that emerges in the lower-temperature SC regime below $T^*$.

**Logarithmically slow resistance relaxation**

To further probe the zero-field TRS-breaking state, we monitor the temporal evolution of the sheet resistance $R_S(\tau)$ after a magnetic-field perturbation, in which an external field is applied and then rapidly removed. As schematically illustrated in Fig. 3**a**, an external magnetic field of $B_\perp$ = 2 T is first applied and then rapidly removed. The relaxation measurement begins at $\tau$ = 0 once the field reaches zero. Figures 3**b**-**g** show representative $R_S(\tau)$ curves for the $La_{1.95}Pr_{1.05}Ni_2O_7$ thin film below 35 K, corresponding to the lower-temperature portion of SC transition regime below $T^*$. The $R_S(\tau)$ curves exhibit a clear logarithmically slow decrease with time, with no sign of saturation over the 20-min measurement window. Such resistance relaxation even persists over 215 minutes in a $La_{2.85}Pr_{0.15}Ni_2O_7$ film (Extended Data Fig. 5). The logarithmically slow relaxation is a hallmark of glassy dynamics in glasses and complex systems[50-52]. Further evidence for glassiness is provided by the waiting-time $\tau_w$ dependence of the relaxation and the corresponding $\tau_w$-scaling, unique characteristic signatures of aging in a glassy state[53] (Extended Data Fig. 6). In addition, the magnitude of the $R_S(\tau)$ relaxation increases when the perturbing field is removed more rapidly (Extended Data Fig. 2**a**), reminiscent of the dependence of glassy relaxation on the rate at which the system is driven into the nonequilibrium state. By contrast, $R_S(\tau)$ measured using the same protocol above 36 K remains essentially time-independent, indicating that the glassy dynamics are confined to the lower-temperature SC transition regime below $T^*$, and are absent both in the higher-temperature portion of the SC transition and in the normal state above $T_c^{onset}$.

The coefficient of the time-dependence is extracted by fitting with $R_S \propto -\beta \log_{10} \tau$ (representative fits shown in Fig. 3**b-g**). Figure 3**k** shows the temperature evolution of coefficient $\beta$, which remains finite below 35 K and vanishes at higher temperatures. These measurements define a characteristic glassy-dynamics temperature, $T_g$ = 35 K, which closely matches $T^*$ and lies near the hysteresis onset temperature $T_{hys}$. Moreover, once the perturbing field exceeds the saturation field of hysteresis, the subsequent resistance relaxation becomes insensitive to further increases in field, indicating that the hysteretic and glassy characteristics share a common characteristic magnetic-field scale (Supplementary Fig. S39). Together, these observations establish a close correspondence between glassiness and the TRS-breaking characteristics, both of which emerge in the lower-temperature portion of the SC transition regime below $T^*$.

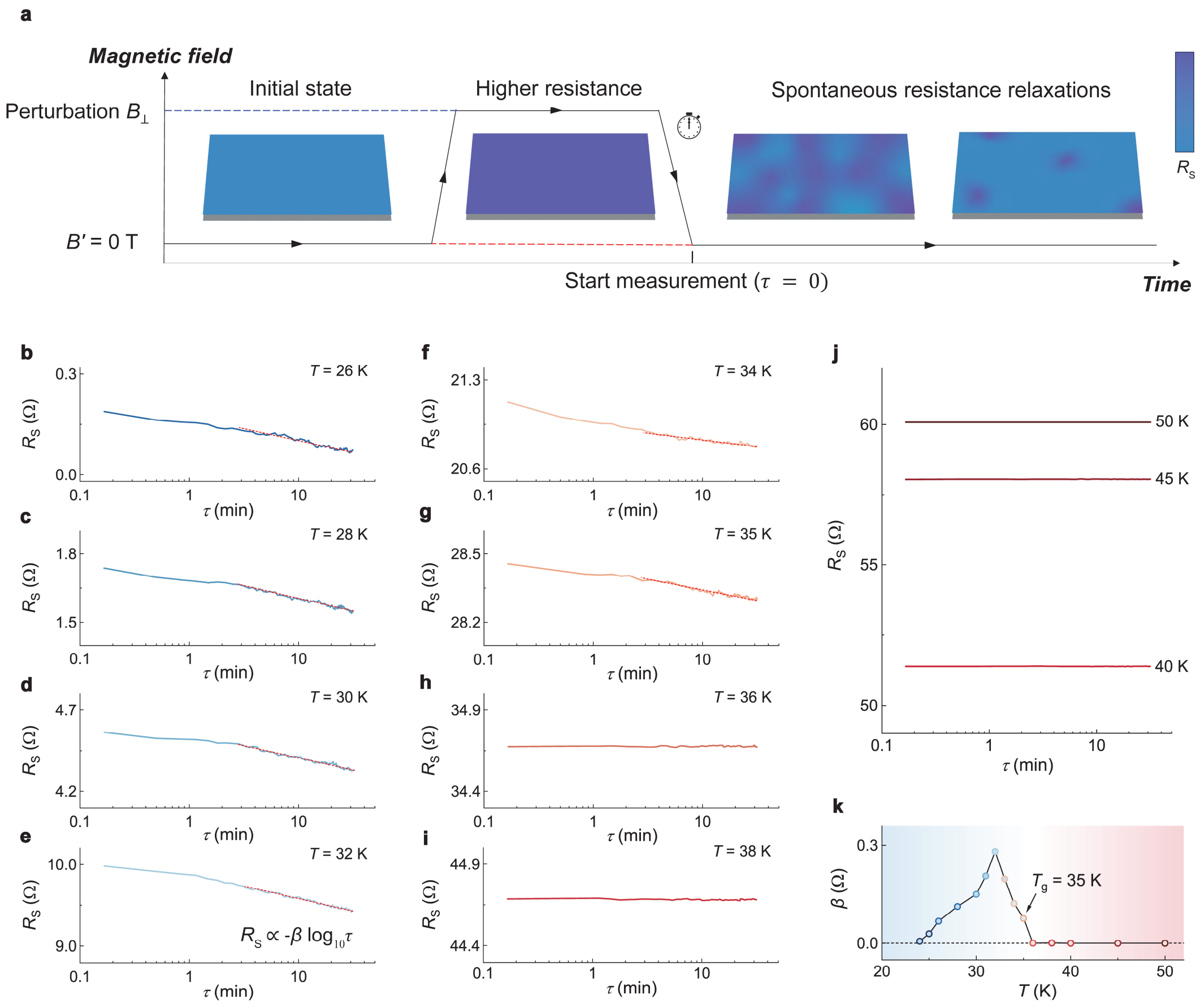

**Fig. 3 | Time-dependent resistance relaxations in $La_{1.95}Pr_{1.05}Ni_2O_7$ thin film (S1). a,** Schematic illustration of the measurement process. At a fixed temperature, an external $B_\perp$, exceeding the saturation field of the hysteresis, is applied. The applied magnetic field partially suppresses superconductivity, thereby driving the initial state into a higher-resistance state. Then, the external magnetic field is rapidly reduced to zero, leading to a perturbed zero-field state. Upon reaching zero field, the time-dependent

resistance measurements are initiated at the counting time $\tau = 0$. **b** to **g,** Time-dependent sheet resistance $R_S(\tau)$ below 35 K, showing nearly logarithmic time-dependent resistance relaxations. **h** to **j,** $R_S(\tau)$ above 36 K, showing no discernible time-dependent variation. **k,** Temperature-dependence of the relaxation coefficient $\beta$, extracted from fits to $R_S \propto -\beta \log_{10} \tau$. Red dashed lines from **b** to **g** are the fit curves to extract $\beta$. The glassy dynamics emerge below $T_g = 35$ K.

**Phase diagram and Temperature hierarchy**

To clarify how TRS-breaking and glassy phenomena relate to superconductivity, we construct an oxygen concentration-dependent phase diagram and establish the hierarchy of characteristic temperatures (Fig. 4**c** and **d**). The superconductivity in the bilayer nickelate film is highly sensitive to oxygen content[28-30], offering a practical tuning parameter. By employing *in situ* reduction of oxygen content (See **Methods**), we characterize the SC properties and hysteretic characteristics across a wide range of oxygen concentration in the $La_{1.95}Pr_{1.05}Ni_2O_7$ film S1. Here, the effective oxygen content $p$ is determined by the empirical relation $p = 143.7\ \Omega / R_S(T = 200\ K)$ [54,55], where $p = 1$ is assigned to the initial state with $R_S(T = 200\ K) = 143.7\ \Omega$.

With progressive oxygen reduction, the film becomes increasingly resistive, the superconducting transition gradually broadens (Fig. 4**a**), and, notably, $T_c^{zero}$ is systematically suppressed (inset of Fig. 4**a**). The magnetoresistance hysteresis, measured at 26 K, shrinks with decreasing $p$ (Fig. 4**b**), as illustrated by the decreasing hysteresis width $\Delta B_{max}$ (marked by arrows in Fig. 4**b**). Summarizing these characteristic quantities of superconductivity and hysteresis, the $p$-$T$ phase diagram is constructed in Fig. 4**c**. Across the investigated $p$ range, hysteresis consistently emerges just within the SC transition regime, remaining well below $T_c^{onset}$. Overall, $T_{hys}$, $\Delta B_{max}$, and $T_c^{zero}$ decrease systematically with decreasing $p$. This simultaneous suppression reveals a positive correlation between the emergence of hysteresis and the establishment of a globally phase coherent, zero-resistance SC state, suggesting a cooperative rather than competitive relationship among them.

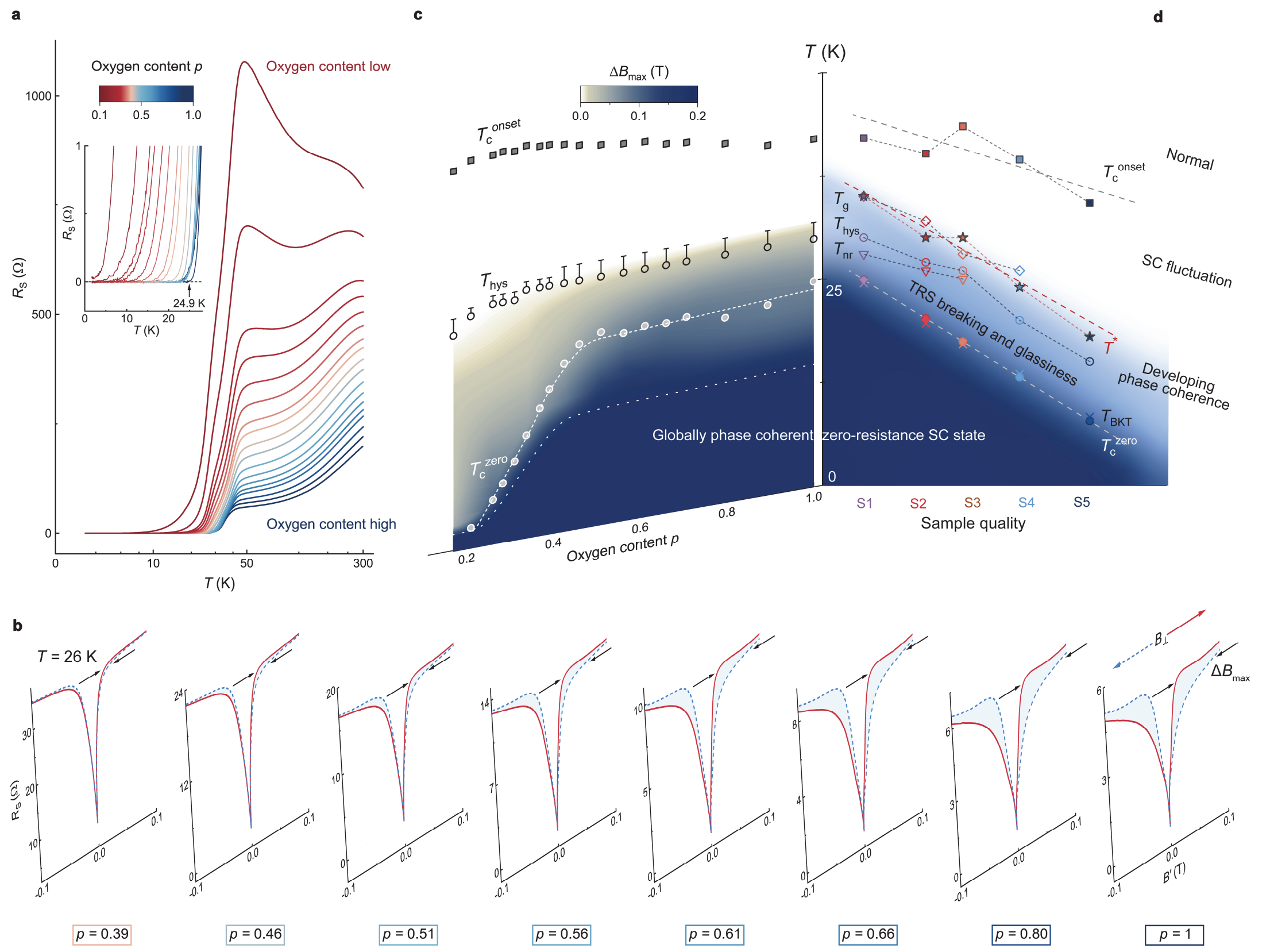

**Fig. 4 | Oxygen-concentration dependence in $La_{1.95}Pr_{1.05}Ni_2O_7$ thin film (S1) and hierarchy of characteristic temperatures across multiple films. a,** $R_S(T)$ curves at different oxygen contents from $p = 1$ to $p = 0.17$ in $La_{1.95}Pr_{1.05}Ni_2O_7$ thin film S1. Here, a shifted log scale $x' = \log_{10}(x+3)$ is utilized for visual clarity. Effective oxygen content $p$ is determined by $p = 143.7\ \Omega / R_S(T = 200\ \mathrm{K})$. **b,** Magnetoresistance at 26 K under different oxygen contents. Blue areas highlight the hysteresis. Black arrows schematically mark the hysteresis width $\Delta B_{max}$. The magnetic fields are calibrated by the Hall sensor (denoted as $B'$). **c,** Phase diagram as a function of temperature $T$ and oxygen content $p$, incorporating $T_c^{onset}$ (onset of SC transition), $T_c^{zero}$ (zero-resistance state) and $T_{hys}$ (onset of hysteresis). Error bars represent the temperature intervals during the measurements. The color scale represents the extracted $\Delta B_{max}$. White dotted line demarcating the dark blue region represents the lowest limits of our resistance measurement, imposed by the vanishingly small resistance signal or the base instrumental temperature. **d,** Hierarchy of characteristic temperatures associated with superconductivity, TRS breaking, and glassy phenomena across different (La, Pr, Sm)$_3$Ni$_2$O$_7$ films. $T_c^{onset}$, $T^*$, $T_c^{zero}$ (solid symbols), and $T_{BKT}$ delineate the normal state (white), the higher-temperature SC fluctuation regime (white), the lower-temperature SC phase coherence-developing regime (light blue), and the globally phase coherent zero-resistance state (dark blue), respectively. $T_{hys}$, $T_{nr}$, and $T_g$ (hollow symbols) generally lie between $T^*$ and $T_c^{zero}$, placing the TRS-breaking and glassy phenomena within the lower-temperature regime where SC phase rigidity/coherence is developing.

We then compare the characteristic temperatures associated with superconductivity, TRS breaking, and glassiness across all measured (La, Pr, Sm)$_3$Ni$_2$O$_7$ films ($p$ = 1), establishing a systematic temperature hierarchy (Fig. 4**d**). Specifically, $T_c^{onset}$, $T_c^{zero}$, and $T_{BKT}$ (which closely matches $T_c^{zero}$) demarcate the normal state, the SC transition regime, and the globally phase coherent zero-resistance SC state (guided by dashed lines). The characteristic temperature $T^*$, associated with the onset of developing SC phase rigidity, marks a crossover that divides the SC transition into a higher-temperature SC fluctuation-dominated regime ($T > T^*$) and a lower-temperature regime of developing SC phase coherence ($T < T^*$). We further include the onset temperatures of $R_S(B)$ hysteresis ($T_{hys}$), non-reciprocal charge transport ($T_{nr}$), and glassy dynamics ($T_g$), which characterize the emergence of TRS-breaking and glassy phenomena.

Figure 4**d** shows that $T_{hys}$, $T_{nr}$, and $T_g$ generally lie between $T^*$ and $T_c^{zero}$, while remaining well separated from $T_c^{onset}$. This temperature hierarchy indicates that TRS breaking and glassiness emerge specifically in the lower-temperature SC transition regime where finite-scale SC phase coherence/rigidity is developing toward establishing global SC phase coherence, pointing to a common origin in the SC phase sector. Notably, this temperature hierarchy is reproduced across five different films with distinct characteristic energy scales, underscoring its robustness. Moreover, $T_{hys}$, $T_{nr}$, and $T_g$ systematically increase in higher-quality films with higher $T_c^{zero}$, arguing against extrinsic disorder as the primary origin of the observed TRS-breaking and glassy phenomena. Together, these findings strongly suggest that both TRS-breaking and glassy characteristics are intrinsic to the unconventional SC state in (La, Pr, Sm)$_3$Ni$_2$O$_7$ films, potentially originating from the internal phase sector of superconductivity.

**Spontaneous magnetic-flux patterns imaged by sSQUID**

Complementing the electrical transport measurements, we perform the sSQUID measurements to directly probe spontaneous magnetic signatures associated with TRS breaking[56,57] as well as their relationship with superconductivity. The sSQUID measurements are carried out on a La$_{1.35}$Pr$_{1.65}$Ni$_2$O$_7$ film (sample S2) with $T_c^{onset}$ = 40.2 K, $T^*$ = 30 K, and $T_c^{zero}$ = 20.2 K. Consistent with the transport properties discussed above, this sample also exhibits hysteresis below 27 K and non-reciprocity below 26 K (Supplementary Figs. S17-S19).

After zero-field cooling (ZFC) from 50 K to 11.5 K, sSQUID magnetometry images resolve clear spatial magnetic-flux patterns at zero applied field below 20 K (Fig. 5**a** to **c**), which is below $T_c^{zero}$. These magnetometry images reveal the emergence of spontaneous magnetic flux in zero external field, providing direct evidence for spontaneous TRS breaking in the SC regime in La$_{1.35}$Pr$_{1.65}$Ni$_2$O$_7$ film.

Several observations argue against the extrinsic causes of the observed magnetic-flux signals. First, the stripe-like patterns are inconsistent with conventional vortex stray fields, as isolated vortices typically appear as approximately circular, dot-like features in sSQUID images[58,59]. Second, the extracted Pearl penetration length $\Lambda$, which sets the characteristic spatial scale of vortex fields in a thin film, exceeds 200 μm (middle panel in Fig. 5**i**), which is substantially larger than both the scanned area and the dimensions of the observed structures. Third, the spatially organized stripe-like morphology, together with the disappearance of the signals just above $T_c^{zero}$, is incompatible with a simple picture of randomly distributed, static magnetic impurities. Fourth, the zero-field flux patterns cannot be readily attributed to Meissner screening currents associated with SC diamagnetism, since the observed magnetic-flux signals are even stronger at zero applied field than under an out-of-plane field of 1 mT (Supplementary Fig. S24).

These zero-field magnetic-flux patterns progressively weaken upon warming from 14 K to 20 K, and become indiscernible at 26 K (Fig. 5**d**). Simultaneously, the local susceptibility was measured by sSQUID susceptometry, which also shows SC diamagnetism from 14 K to 20 K (Supplementary Fig. S20), but no diamagnetic signal above 26 K. In thin films, the SC diamagnetic susceptibility is proportional to the inverse Pearl penetration length, and is therefore directly related to local superfluid density and SC phase stiffness/rigidity. Thus, this concurrence of spontaneous magnetic flux and SC diamagnetism strongly suggests the intrinsic association between TRS-breaking signatures and superconductivity in $(La, Pr, Sm)_3Ni_2O_7$ films.

Zero-field magnetometry images are measured after field cooling (FC) in +1 mT or −1 mT, with the external field removed at the base temperature before imaging. Compared with the ZFC results, the zero-field magnetometry images after +1 mT FC exhibit more pronounced magnetic-flux patterns with significantly enhanced amplitudes (Fig. 5**e**). Reversing the cooling field to −1 mT essentially reverses the polarity of the zero-field magnetic-flux patterns (Fig. 5**g**), demonstrating a clear field-training effect consistent with the applied field biasing the populations or configurations of underlying TRS-conjugate domains.

Upon warming after −1 mT FC, both the magnetic-flux patterns and SC diamagnetism remain discernible up to approximately 25 K, and become indiscernible above 26 K (Supplementary Figs. S22 and S23). Figure 5**i** directly compares the temperature dependences of the spatially averaged flux magnitude $|\Phi|$ and the susceptibility $\chi'$ in zero field. Furthermore, the Pearl penetration length $\Lambda$ is extracted from sSQUID approach measurements at a representative location[60] (Supplementary Figs. S26 and S27), from which the corresponding local phase stiffness $J_s$ is determined (middle panel in Fig. 5**i**). For comparison, the hysteresis width $\Delta B_{max}$ and the zero-field non-reciprocal critical current $\Delta I_c$ measured on the same film are shown in the bottom panel in Fig. 5**i**. Together, these results consistently show that TRS-breaking signatures ($\Phi$, $\Delta B_{max}$, and $\Delta I_c$), SC diamagnetic susceptibility $\chi'$, and local SC phase stiffness $J_s$ all emerge within a common temperature range, highlighted by the white region in Fig. 5**i**. This concurrence directly links the onset of TRS breaking to the local development of SC phase stiffness/rigidity, supporting a common origin in the phase sector of superconductivity. Importantly, no magnetic-flux signal is observed once the local SC diamagnetic response disappears, further arguing against extrinsic magnetic origins such as conventional static magnetism.

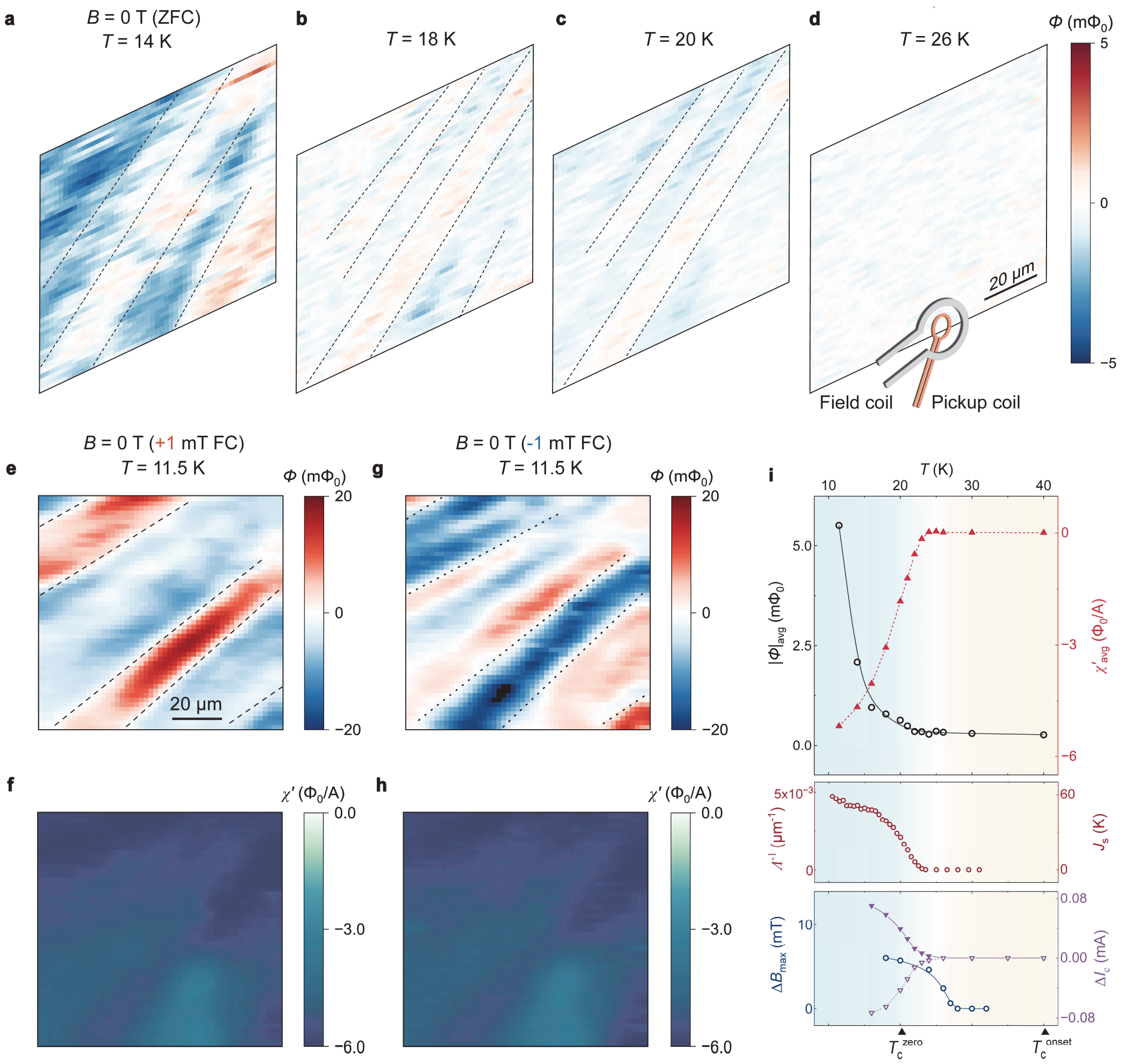

**Fig. 5 | Scanning SQUID images of $La_{1.35}Pr_{1.65}Ni_2O_7$ film (S2) at zero magnetic field. a-d**, sSQUID magnetometry images measured at 14 K (**a**), 18 K (**b**), 20 K (**c**) and 26 K (**d**) after zero-field cooling (ZFC) from 50 K to 11.5 K. Schematic of the sSQUID coils in **d** is enlarged twofold for clarity. **e-h**, Zero-field magnetometry (**e**, **g**) and corresponding susceptometry (**f**, **h**) images after field cooling (FC). The film was field-cooled from 50 K to 11.5 K in $B_\perp$ = +1 mT (**e**, **f**) or −1 mT (**g**, **h**), with the field removed at 11.5 K. All images are scanned over an area of 90 × 90 μm$^2$. Environmental

background signals, acquired from distant off-sample images, have been subtracted. Dashed and dotted lines are guides to the eye. **i**, Temperature dependence of the sSQUID and electrical transport quantities. Top: magnetic-flux magnitude $|\Phi|$ (black circles) and susceptibility $\chi'$(red triangles), both spatially averaged over the scanned area and measured at zero field after −1 mT FC. Middle: inverse Pearl penetration length $\Lambda^{-1}$ (dark red circles, left axis) and corresponding local SC phase stiffness $J_s$ (right axis). Bottom: hysteresis width $\Delta B_{max}$ (blue circles) and zero-field non-reciprocal critical current $\Delta I_c$ (purple triangles). $T_c^{onset}$ and $T_c^{zero}$ are marked on the temperature axis for reference. Colored regions are guides to the eye, and the intervening white region highlights the concomitant onsets of TRS-breaking signatures and local SC phase stiffness.

**Discussion**

Before considering the possible microscopic origin of the observed TRS-breaking and glassy phenomena, we first examine several potential extrinsic mechanisms. First, our key experimental observations are consistently reproduced in multiple films with different compositions and characteristic temperature scales, indicating that these phenomena are robust and potentially generic to the R-P bilayer nickelate films. Moreover, the characteristic temperatures associated with TRS breaking and glassy dynamics systematically increase in higher-quality films with enhanced superconductivity and reduced disorder (Fig. 4**d** and Extended Data Fig. 1), arguing strongly against extrinsic inhomogeneity or disorder as the underlying physical origin of these phenomena.

Second, instrumental artifacts, including remnant magnetic fields of the magnet, can be excluded. The magnetic field is independently calibrated using a high-resolution InAs Hall sensor, and no hysteresis is observed in the higher-temperature portion of the SC transition regime or in the normal state in all measured $(La, Pr, Sm)_3Ni_2O_7$ films and in a conventional superconductor Nb film (Extended Data Fig. 8). Third, a simple magnetic-impurity scenario is unlikely. All relevant device components are nominally non-magnetic, including the $SrLaAlO_4$ substrate, Pt electrodes, and Al bonding wires, while the hysteresis and nonreciprocity vanish at higher temperatures. In addition, sSQUID magnetometry reveals no signatures characteristic of static magnetic impurities (Fig. 5), and atomic-resolution STEM detects no discernible cation impurities, including magnetic Ni impurities (Fig. 1**b** and **c**). Fourth, a conventional trapped-vortices scenario is disfavored. The hysteresis persists under in-plane magnetic fields, for which the generation of conventional orbital vortices is strongly suppressed by the quasi-two-dimensional geometry (Extended Data Figs. 3 and 4). Moreover, the $R_S(B)$ curves measured under opposite field-sweep directions exhibit minima that coalesce at $B$ = 0, as independently calibrated by the Hall sensor. This behavior fundamentally argues against the scenario of trapped-vortices[61], for which oppositely swept curves are expected to exhibit two split minima at finite and approximately symmetric fields owing to compensation between trapped flux and the applied magnetic field. Additional considerations excluding extrinsic origins are provided in **Methods**.

Our observations provide compelling evidence for spontaneous TRS breaking and intrinsic glassiness in the SC state in nickelate (La, Pr, Sm)$_3$Ni$_2$O$_7$ thin films. The close correspondence of both phenomena with SC phase rigidity points to the phase sector of superconductivity as a possible common origin. A natural candidate framework is a TRS-breaking multicomponent SC phase glass, in which internal relative-phase textures develop within a multicomponent SC order parameter while intrinsically coupled to the overall SC phase. In such a scenario, the relative-phase sector provides an internal TRS-breaking chiral degree of freedom, while its spatial textures undergo slow, glassy freezing alongside the development of the overall SC phase rigidity and coherence below $T^*$.

The bilayer nickelates are multiband systems with multiple Fermi surfaces in the normal state. Multiple nearly degenerate pairing channels have been proposed[10,11], providing a natural and anticipated setting for a multicomponent SC order parameter. Writing

$$\Delta_a(\mathbf{r}) = |\Delta_a(\mathbf{r})|e^{i\theta_a(\mathbf{r})},\ a = 1,2,...,N,$$

the phase degrees of freedom can be separated conceptually into an overall SC phase $\Theta$, and gauge-invariant relative phases

$$\phi_{ab} = \theta_a - \theta_b.$$

The overall phase controls the $U(1)$ phase stiffness and the BKT physics, whereas the relative phases constitute internal degrees of freedom that can support additional discrete ($Z_2$) symmetry breakings[62-64]. In particular, when the free-energy minimum occurs at a relative phase $\phi_{ab} = \pm\phi_0$, with $\phi_0 \neq (0, \pi)$, the two configurations are related by time reversal and carry opposite chirality. Spontaneous selection between these time-reversal-conjugate configurations therefore breaks TRS. Frustrated couplings among multiple SC components can further generate competing relative-phase configurations and spatially textured chiral domains. Importantly, these relative-phase textures are generally coupled to the overall $U(1)$ phase through domain walls, phase gradients, and vortex or fractional-vortex defects. The resulting composite phase texture can possess a large manifold of metastable configurations separated by collective energy barriers, providing a natural route to glassy dynamics. Within this picture, the frozen variables are primarily SC phase configurations rather than microscopic magnetic moments, making the resulting state analogous to a phase or chiral glass. The logarithmically slow relaxation, aging, and field-history dependence observed experimentally can then arise from slow rearrangements among these metastable SC phase textures. Such glassiness is intrinsically generated by frustration in the SC phase sector, although weak structural disorder may still influence domain pinning and relaxation kinetics. The intrinsic glassiness is distinct from conventional disorder-driven spin glass, and behaves differently from glassy magnetism that coexists with superconductivity[65-67].

The development of the internal relative-phase texture and the growth of the overall phase rigidity need not occur independently. Instead, the two sectors are inherently coupled and evolve cooperatively as the system is cooled through the SC transition regime. This provides a unified interpretation of the experimentally observed temperature hierarchy: TRS-breaking and glassy signatures emerge below $T^*$, where finite-scale SC phase rigidity begins to develop, while global coherence and the zero-resistance state are established only at lower temperature near $T_{BKT}$. The same framework naturally accounts for the local concurrence of spontaneous magnetic flux and SC diamagnetism observed by sSQUID, since spatial textures of the internal relative phase can generate spontaneous supercurrents once sufficient local phase rigidity develops. The systematic evolution with sample quality and oxygen concentration is also compatible with this picture, as the TRS-breaking and glassy phenomena track the common characteristic temperature scale associated with the SC phase rigidity.

In summary, our electrical transport and sSQUID measurements demonstrate spontaneous TRS breaking and electronic glassiness intrinsic to the SC order in nickelate (La, Pr, Sm)$_3$Ni$_2$O$_7$ thin films. The emergence of these phenomena tracks the development of SC phase coherence, pointing to a common origin in the internal phase sector of superconductivity. Our observations are naturally consistent with a SC phase glass, in which the intercomponent relative phases of a multicomponent SC order parameter carry TRS-breaking chirality and freeze into a textured phase glass. Although the microscopic origin remains to be established, our findings uncover a previously unrecognized form of SC state, realized in the nickelate films through the coexistence of spontaneous TRS breaking and electronic glassiness. This discovery adds a new dimension to the emergent physics of nickelate superconductors and calls for further experimental and theoretical investigation of its microscopic origin and collective dynamics. More broadly, our findings highlight internal SC phase degrees of freedom as a fertile arena for intertwined symmetry breaking, electronic glassiness, and unconventional high-$T_c$ superconductivity.

## Methods.

### Thin-film synthesis.

$La_{3-x-y}Pr_xSm_yNi_2O_7$ thin films were deposited on $SrLaAlO_4$ (001) substrates (MTI-Kejing) via the gigantic-oxidative atomic-layer-by-layer epitaxy (GAE) technique[32]. Four compositions were investigated in this study: (i) $La_{1.95}Pr_{1.05}Ni_2O_7$ thin film ($x$ = 1.05, y = 0), (ii) $La_{1.35}Pr_{1.65}Ni_2O_7$ thin film ($x$ = 1.65, y = 0), (iii) $La_{2.85}Pr_{0.15}Ni_2O_7$ thin film ($x$ = 0.15, y = 0) and (iv) $La_{2.46}Pr_{0.24}Sm_{0.3}Ni_2O_7$ thin film ($x$ = 0.24, $y$ = 0.3). The $La_{0.95}Pr_{0.05}O_x$, $La_{0.65}Pr_{0.35}O_x$ and $La_{0.82}Pr_{0.08}Sm_{0.1}O_x$ targets were prepared by stoichiometric mixtures of LaO, PrO and SmO powders at 1050 °C for 2 h. The growth temperature and laser fluence for $La_{2.85}Pr_{0.15}Ni_2O_7$ and $La_{2.46}Pr_{0.24}Sm_{0.3}Ni_2O_7$ were 750 °C and 2.0 J/cm$^2$, and the growth parameters for $La_{1.95}Pr_{1.05}Ni_2O_7$ were 850 °C and 2.1 J/cm$^2$. The deposition atmosphere consists of two components: purified ozone and oxygen. After growth, the $La_{2.85}Pr_{0.15}Ni_2O_7$ film was post-annealed in an oxidation chamber equipped with an ambient-pressure ozone generator and *in-situ* resistance monitoring; whereas the $La_{1.95}Pr_{1.05}Ni_2O_7$, $La_{1.35}Pr_{1.65}Ni_2O_7$ and $La_{2.46}Pr_{0.24}Sm_{0.3}Ni_2O_7$ films were superconducting without post annealing. For the $La_{1.95}Pr_{1.05}Ni_2O_7$, $La_{1.35}Pr_{1.65}Ni_2O_7$ and $La_{2.46}Pr_{0.24}Sm_{0.3}Ni_2O_7$ films, an insulating (La, Pr, Sm)O-$NiO_2$-(La, Pr, Sm)O buffer layer was grown between the thin film and the as-received $SrLaAlO_4$ (001) substrate to mitigate the interfacial structural discontinuity.

The X-ray diffraction (XRD) patterns (Supplementary Fig. S1**a**) reveal consecutive Bragg peaks along the out-of-plane axis without indication of impurity phases, confirming periodicity and crystallinity within only 3 unit cells. Reflection high-energy electron diffraction (RHEED) oscillations (Supplementary Fig. S1**b**) reveal that successive atomic layers were deposited layer by layer onto the $SrLaAlO_4$ (001) substrate by laser ablation of the respective targets, reflecting the stoichiometry and coverage of the film. The stable oscillation amplitude indicates highly reproducible growth kinetics for each monolayer and precise control of stoichiometry.

Standard Hall bar electrodes with Pt were deposited by magnetron sputtering with a shadow mask. The electrodes were bonded to sample holders with Al wires. The contact resistance of the electrodes on $La_{1.95}Pr_{1.05}Ni_2O_7$ films is smaller than 2 Ω at 2 K.

### Transport measurements

Transport measurements were performed in two commercial physical property measurement systems from Quantum Design: a Re-liquefier system with a 16 T-magnet (PPMS-16 T) and a cryogen-free system with a 14 T-magnet (PPMS Dynacool-14 T). The PPMS-16 T is equipped with a rotator with an accuracy of 0.053°, and the PPMS Dynacool-14 T is equipped with a high-resolution rotator with an accuracy of 0.011°.

For the *I*-*V* curves and the controlled measurements with d.c. current, an AC/DC current source meter (Keithley 6221) was used to apply a d.c. current, and a nanovoltmeter (Keithley 2182 A) was used to measure the voltage. The applied current is cut off during the intervals of changing current, in order to eliminate the heating effect. For the controlled measurements with a.c. current, an AC/DC current source meter (Keithley 6221) was used to apply an a.c. current, and lock-in amplifiers (Stanford Research Systems SR830) were used to measure voltage.

Two cryogenic transverse InAs Hall sensors (Lakeshore HGCT-3020) were used to calibrate the values of the applied magnetic fields.

**Oxygen content control**

The oxygen content control follows the previously reported methodology[68]. In the vacuum environment (~10 Torr helium gas), the progressive resistance increase was observed above 200 K, which is attributed to the spontaneous oxygen loss from the films. To systematically control the oxygen concentration, we hold the samples above 300 K in the PPMS under vacuum, while monitoring the increasing resistance. Upon reaching the targeted oxygen concentration, with resistance serving as an indicator, the films were cooled down for measurement.

**Scanning SQUID microscopy**

The sSQUID measurements were carried out in a cryostat with a base temperature of 10 K and a maximum out-of-plane magnetic field of 1 mT. The sensors were fabricated in the SQUID-on-chip configurations. The diameter of the pickup coil is 3 μm and that of the field coil is 14 μm. The nano-SQUID chip was thermally isolated from the variable-temperature sample stage and maintained at a working temperature of approximately 4.50 K. For the scanning SQUID measurements, two different modes were employed and were performed simultaneously: (i) Magnetometry ($\Phi$) measures the flux through the pickup loop. (ii) Susceptometry ($\chi$) measures the inductive response of the film through the pickup loop to an external modulation a.c. current through the field coil ($I_F$), where $\chi = d\Phi/dI_F$. $I_F$ = 130 μA with a frequency of 1.1 kHz was applied. The environmental background signals were acquired from distant off-sample images, and were then subtracted.

**Anisotropic hysteresis under different magnetic field orientations**

We observe the anisotropic magnetoresistance hysteresis under different magnetic field orientations in $La_{2.46}Pr_{0.24}Sm_{0.3}Ni_2O_7$ thin film (S4). Under opposite swept $B_\perp$, this $La_{2.46}Pr_{0.24}Sm_{0.3}Ni_2O_7$ film ($T_c^{onset}$ = 39.5 K, $T_c^{zero}$ = 13.1 K) shows qualitatively consistent magnetoresistance hysteresis with coalescing minima below 20 K (Extended Data Fig. 3). The hysteresis is also observed under an in-plane magnetic field $B_{//}$ in the lower-temperature portion of SC transition regime, which disappears above 22 K (Extended Data Fig. 3**b**). Below 18 K, the hysteresis loop seems to be unsaturated up to 15 T (the maximum applied field), suggesting a substantially larger saturation field along in-plane orientation than out-of-plane one. Considering the absence of $B_{//}$-induced vortices due to confined dimensionality in $La_{2.46}Pr_{0.24}Sm_{0.3}Ni_2O_7$ thin films, the hysteresis under $B_{//}$ cannot be attributed to the trapped vortices. Therefore, the emergence of hysteresis in the lower-temperature SC transition regime under both $B_{\perp}$ and $B_{//}$ supports the intrinsic nature of TRS breaking in SC state.

To characterize the magnetic field orientation-dependence, the magnetoresistances are measured under the opposite sweeping magnetic fields $B_\theta$ at different polar angles $\theta$ from 0.00° (out-of-plane, Extended Data Fig. 3**c**) to 90.00° (in-plane, Extended Data Fig. 3**i**), and then to higher degrees (Extended Data Fig. 3**j**). As $\theta$ increases from 0.00° to 88.29°, the hysteresis loop expands and the saturation field increases. This trend may be associated with the diminished $B_{\perp}$-component that is proportional to $\cos\theta$. However, when $\theta$ further increases, the hysteresis loop undergoes an anomalous evolution: it significantly shrinks and nearly closes at 89.89°, and then fully re-opens around 90.00° with a reversed polarity. This abrupt change with polarity reversal within a narrow $\theta$ range around 90.00° (in-plane) cannot be attributed to the effect of a $B_{\perp}$-component.

We further quantify the hysteresis amplitude by extracting the vertical displacements of hysteresis, defined as $\Delta R_{S,\,max}$ (Extended Data Fig. 3**k**). Specifically, $\Delta R_{S,\,max}(\theta)$ is defined as the maximum magnitude of $\Delta R_S(B) = R_S^{\uparrow}(B) - R_S^{\downarrow}(B)$ at different $\theta$, and its sign is determined by $\Delta R_S(B > 0)$. Here, ↑ denotes magnetic fields from −15 T to 15 T, and ↓ in reverse. With increasing $\theta$ from 0° to approximately 88°, $\Delta R_{S,\,max}$ remains a positive and nearly invariant value around 0.22 Ω, showing a weak $\theta$-dependence. As $\theta$ further increases from 88° to 90°, $\Delta R_{S,\,max}$ decreases rapidly from positive 0.22 Ω (red diamonds), goes through zero (0 Ω marked by black dashed line), and then reaches negative 0.29 Ω (blue diamond). With increasing $\theta$ from 90° to 180°, $\Delta R_{S,\,max}$ exhibits a mirror symmetric $\theta$-dependence about the in-plane orientation. The overall $\theta$-dependent $\Delta R_{S,\,max}$ is centered at $\theta = 90°$, confirming the intrinsic nature of the rapid change, which cannot be ascribed to misalignment of $\theta$. The singular $\theta$-dependence confined within a small angle is reminiscent of the interlayer Josephson coupling in layered superconductors[69] under $B_{//}$. Further investigations focusing on this confined $\theta$-range are required to elucidate the underlying microscopic origin.

Similar singular $\theta$-dependence is reproduced in the $La_{2.85}Pr_{0.15}Ni_2O_7$ film (S5, Extended Data Fig. 4 measured in PPMS-16 T, and Supplementary Fig. S52 in PPMS Dynacool-14 T).

**Exclusion of trapped vortices for magnetoresistance hysteresis**

Several observations argue against conventional trapped vortices explanations for the observed magnetoresistance hysteresis. First of all, hysteresis is observed under in-plane magnetic fields $B_{//}$, despite the absence of field-induced vortices due to the confined dimensionality. Second, no hysteretic behavior is observed in the higher-temperature regime of superconducting transition, where the vortices and pinning sites should still exist. Third, several key features of the observed hysteresis explicitly contradict the behaviors of trapped vortices at grain boundaries in granular films (referred to as two-level critical-state model) [61]. From the perspective of the material systems, our films are pure-phase single-crystalline[29], instead of the inhomogeneous granular films. From the perspective of hysteresis behaviors, our $R_S(B)$ curves under opposite swept magnetic fields exhibit coincident minima at 0 T (confirmed by the Hall sensor). This behavior is fundamentally against the trapped vortices scenario, where two sharply split minima emerge at symmetric magnetic fields, resultant from the cancellation between the trapped flux and the applied magnetic field. Additionally, the minima at 0 T remain identical regardless of the maximum applied field (Supplementary Fig. S45), inconsistent with the trapped vortices model where resistance minima shift to higher field as maximum applied fields increase. Fourth, magnetoresistance hysteresis is insensitive to measurement current amplitudes (Extended Data Fig. 2), switching between a.c. or d.c. currents (Supplementary Fig. S50), or changes in a.c. frequencies (Fig. S46). This current-independence indicates that the vortex dynamics is not dominant, since a larger driving current would induce a stronger Lorentz force to de-pin the vortices.

**Exclusion of trapped vortices for glassy dynamics**

First, the glassy dynamics cannot be attributed to a vortex glass state, as such a scenario is incompatible with the observed $\tau_w$-scaling behavior[53] (Extended Data Fig. 6). Second, the resistance relaxation could be triggered by a perturbation of in-plane magnetic fields (Extended Data Fig. 6), where field-induced vortices should be absent. Third, the glassy dynamics are insensitive to measurement current (Extended Data Fig. 2).

**Exclusion of the extrinsic mechanisms for the TRS breaking**

First of all, it should be emphasized that the magnetoresistance hysteresis is consistently reproduced in five different samples with varying compositions: two $La_{1.95}Pr_{1.05}Ni_2O_7$ thin films (S1 and S3), a $La_{1.35}Pr_{1.65}Ni_2O_7$ film (S2), a $La_{2.46}Pr_{0.24}Sm_{0.3}Ni_2O_7$ film (S4) and a $La_{2.85}Pr_{0.15}Ni_2O_7$ film (S5). The comparative analysis across these films reveals that the TRS breaking systematically emerges at higher temperatures in higher-quality films with enhanced superconductivity and reduced disorder strength (Fig. 4, Extended Data Fig. 1), which strongly argues against the extrinsic inhomogeneity-related origin, and demonstrates the intrinsic nature. Second, we rule out the issue of the remnant magnetic field in our measurement system. The observed temperature- and magnetic field orientation-dependence of the hysteresis loops cannot be attributed to the possible remnant field of the system, since the remnant field should not vary with temperatures and rotation angles of the sample. Moreover, the magnetic fields are calibrated by an InAs Hall sensor (Fig. 2). Furthermore, we measured a conventional superconductor Nb thin film in the same equipment, and the $R_S(B)$ curves show no distinguishable hysteresis feature (Extended Data Fig. 8). These results unambiguously exclude the possible artifacts related to the magnet in the measurement system.

Our results cannot be explained by the extrinsic magnetism from magnetic impurities. First, no magnetic-flux signal associated with magnetic impurity is detected in the sSQUID magnetometry (Fig. 5), and no discernible cation impurities such as magnetic Ni are detected in the atomic-resolution STEM images (Fig. 1**b** and **c**). Second, the hysteresis is only observed in the lower-temperature SC regime, with no detectable signature in the higher-temperature regime (Fig. 2). The comparative trend and progressive oxygen concentration-modulation also disfavor the extrinsic impurity-related origin. Our observations are also incompatible with the scenario where magnetism arises at low temperatures via Ruderman–Kittel–Kasuya–Yosida (RKKY) interaction among the potential magnetic impurities. In this RKKY framework, the transition temperature is governed by the coupling strength and the carrier density, and thereby the transition to spin-glass state should emerge at higher temperatures as carrier density increases[70]. In our (La, Pr, Sm)$_3$Ni$_2$O$_7$ films, the carriers are electron-type, and the carrier density increases with reduced oxygen content (loss of $O^{2-}$). The oxygen concentration-modulation shows that the onset temperature of the TRS breaking SC state decreases with the increasing carrier density, which contradicts the RKKY scenarios.

The TRS breaking, evidenced by magnetoresistance hysteresis and non-reciprocity with two polarities, is not due to the glassy dynamics that leads to resistance relaxations. As shown in Extended Data Fig. 2, the hysteresis remains nearly invariant with varying slow sweep rates. With slowly varied magnetic field where no discernible resistance relaxation is detected, the non-reciprocity with two polarities is observed in a static condition (Supplementary Fig. S14). These results collectively indicate that the TRS breaking is not due to the dynamics of magnetic remanence.

**Data availability**

All data needed to evaluate the conclusions in the study are present in the paper and/or the Supplementary Information. The data that support the findings of this study are available from the corresponding author upon request.

**Acknowledgements**

We acknowledge the discussions with Yi-feng Yang, Haiwen Liu, and Ziqiao Wang. This work was financially supported by the National Natural Science Foundation of China [Grant No. 12488201 (J.W.)], the Innovation Program for Quantum Science and Technology [2021ZD0302403 (J.W.)], the Natural Science Foundation of China [Grant No. 92565303 (Z.C., Y.L., H.H., and G.Z.)], the National Key Research and Development Program of China [No. 2024YFA1408101 (Z.C., H.H., and H.W.), No. 2022YFA1403101 (Z.C.), No. 2023YFA1406500 (Y.L.), No. 2022YFA1403103 (Y.L.)], the Natural Science Foundation of China [Grant No. 12374455 (Z.C.), Grant No. 52388201 (Q.-K.X.), Grant No. 12504166 (G.Z.), Grant No. 12504161 (H.W.), Grant No. 12174442 (Y.L.)], Young Elite Scientists Sponsorship Program by CAST [No. 2023QNRC001 (Y.L.)], Young Elite Scientists Sponsorship Program by BAST [No. BYESS2023452 (Y.L.)], the Fundamental Research Funds for the Central Universities and the Research Funds of Renmin University of China [No. 22XNKJ20 (Y.L.)], the Guangdong Provincial Quantum Science Strategic Initiative [Grant No. GDZX2401004 (Z.C.), Grant No. GDZX2201001 (Z.C.)], the Shenzhen Science and Technology Program [Grant No. KQTD20240729102026004 (Z.C.)], and the Shenzhen Municipal Funding Co-Construction Program Project [Grant No. SZZX2301004 (Z.C.), Grant No. SZZX2401001 (Z.C.)], the China Postdoctoral Science Foundation [Grant No. 2024M760063 (H.J), Grant No. GZB20240294 (H.W.), Grant No. 2024M751287 (H.W.)], Z.W. is supported by the U.S. Department of Energy, Basic Energy Sciences Grant DE-FG02-99ER45747, G.-M.Z. acknowledges the support from the National Key Research and Development Program of MOST of China [Grant No. 2023YFA1406400], Z. C. acknowledges the support from International Station of Quantum Materials.

**Author contributions**

J.W. conceived the project. J. W. and Z.C. supervised the project. H.J., Z.X., M.L. and L.P. performed the transport measurements and analyzed the data under the guidance of J.W.. Y.C., G.Z., H.W. and H.H. performed the thin-film growth and sample characterization under the guidance of Q.-K.X. and Z.C. K.L., R.L. and Y.W. performed the sSQUID measurements. J.G. and Y.L. provided the support in data interpretation. M.G. and Y.L. performed the magneto-optic measurements. G.-M.Z. and Z.W. contributed to the theoretical explanations. H.J., Z.W. and J.W. wrote the manuscript with input from all other authors. H.J., Z.X., Y.C. and G.Z. contributed equally to this work.

**Competing interests**

The authors declare no competing interests.

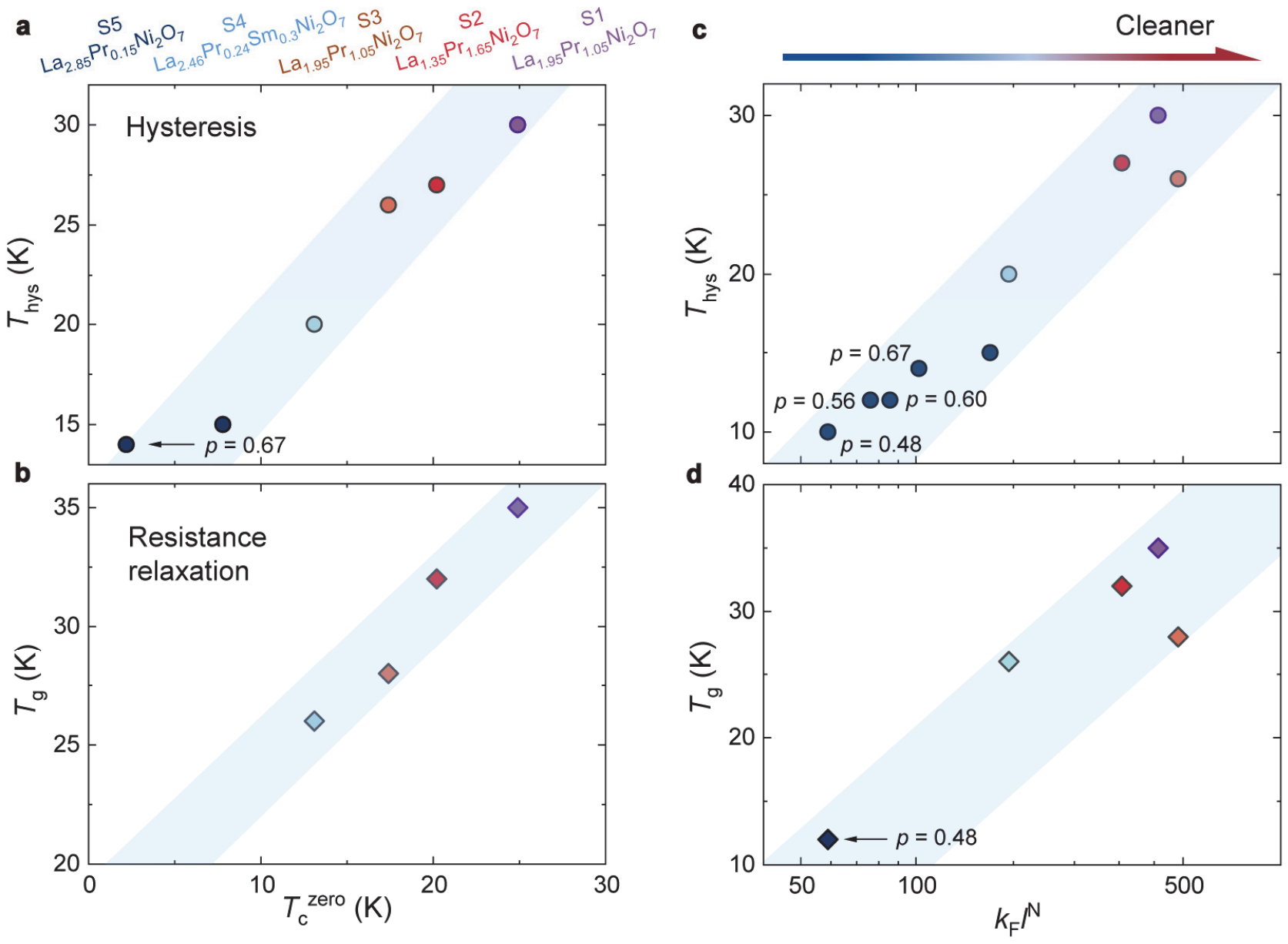

**Extended Data Fig. 1 | Comparative analysis across different (La, Pr, Sm)$_3$Ni$_2$O$_7$ films.** Comparative analysis summarizing the characteristic temperatures for magnetoresistance hysteresis ($T_{hys}$, upper panels) and glassy dynamics ($T_g$, lower panels) as functions of zero-resistance temperatures ($T_c^{zero}$, left columns) or Ioffe-Regel parameters ($k_Fl$, right columns). $k_F$ denotes Fermi wave vector and $l$ represents electron mean free path. The parameter $k_Fl$ provides a measure of the effective disorder strength in a system, where $k_Fl \gg 1$ corresponds to a relatively clean system[71]. The values of $k_Fl$ are estimated[71] by $k_Fl = \frac{h}{e^2}\frac{1}{R_N^{50\ K}}$, where $h$ is Planck's constant, $e$ is elementary charge, and $R_N^{50\ K}$ denotes the normal-state sheet resistance taken at 50 K (The detailed estimation procedure is presented in the Supplementary Information). The positive correlations between $T_{hys}$, $T_g$ and $k_Fl$ argue against a disorder-driven origin, thereby substantiating the intrinsic nature of the TRS breaking hysteresis and glassiness. The characteristic temperatures of the La$_{2.85}$Pr$_{0.15}$Ni$_2$O$_7$ film with different oxygen contents ($p$) are displayed ($p$ = 0.67 to 0.48, as denoted). Unlabeled data points correspond to films at $p$ = 1.

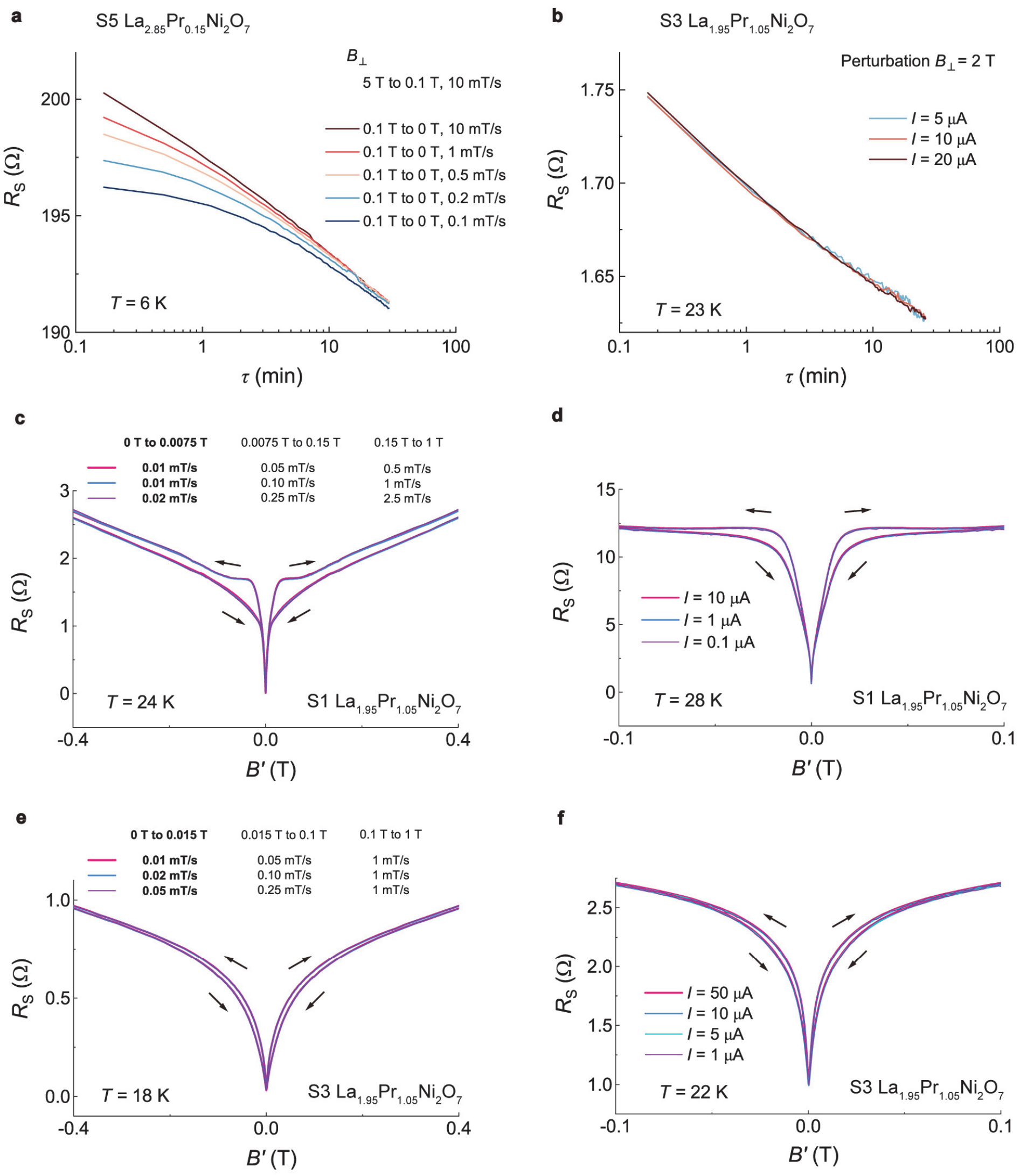

**Extended Data Fig. 2 | Current- and magnetic field-dependence of resistance relaxation and magnetoresistance hysteresis in (La, Pr, Sm)$_3$Ni$_2$O$_7$ thin films. a,** Time-dependent sheet resistance $R_S(\tau)$ curves of the La$_{2.85}$Pr$_{0.15}$Ni$_2$O$_7$ film (S5, $p = 0.48$) measured with different rates, at which the initially applied perturbation magnetic fields are reduced to zero. The $R_S(\tau)$ variations are larger for the faster removal of magnetic fields to zero. **b,** $R_S(\tau)$ curves of the La$_{1.95}$Pr$_{1.05}$Ni$_2$O$_7$ film (S3) measured with different currents, showing no discernible current dependence and thereby disfavoring vortex dynamics. **c** and **e,** Magnetoresistance $R_S(B)$ curves of La$_{1.95}$Pr$_{1.05}$Ni$_2$O$_7$ film (S1, **c**) and La$_{1.95}$Pr$_{1.05}$Ni$_2$O$_7$ film (S3, **e**) measured with different magnetic field-sweep rates, showing no discernible difference. **d** and **f,** Magnetoresistance $R_S(B)$ curves of La$_{1.95}$Pr$_{1.05}$Ni$_2$O$_7$ film (S1, **d**) and La$_{1.95}$Pr$_{1.05}$Ni$_2$O$_7$ film (S3, **f**) measured with different currents, showing no discernible difference at small currents. These results disfavor vortex dynamics as the underlying mechanism. Here, the magnetic fields are calibrated by the Hall sensor, denoted as $B'$. Black arrows indicate the magnetic field-sweep directions.

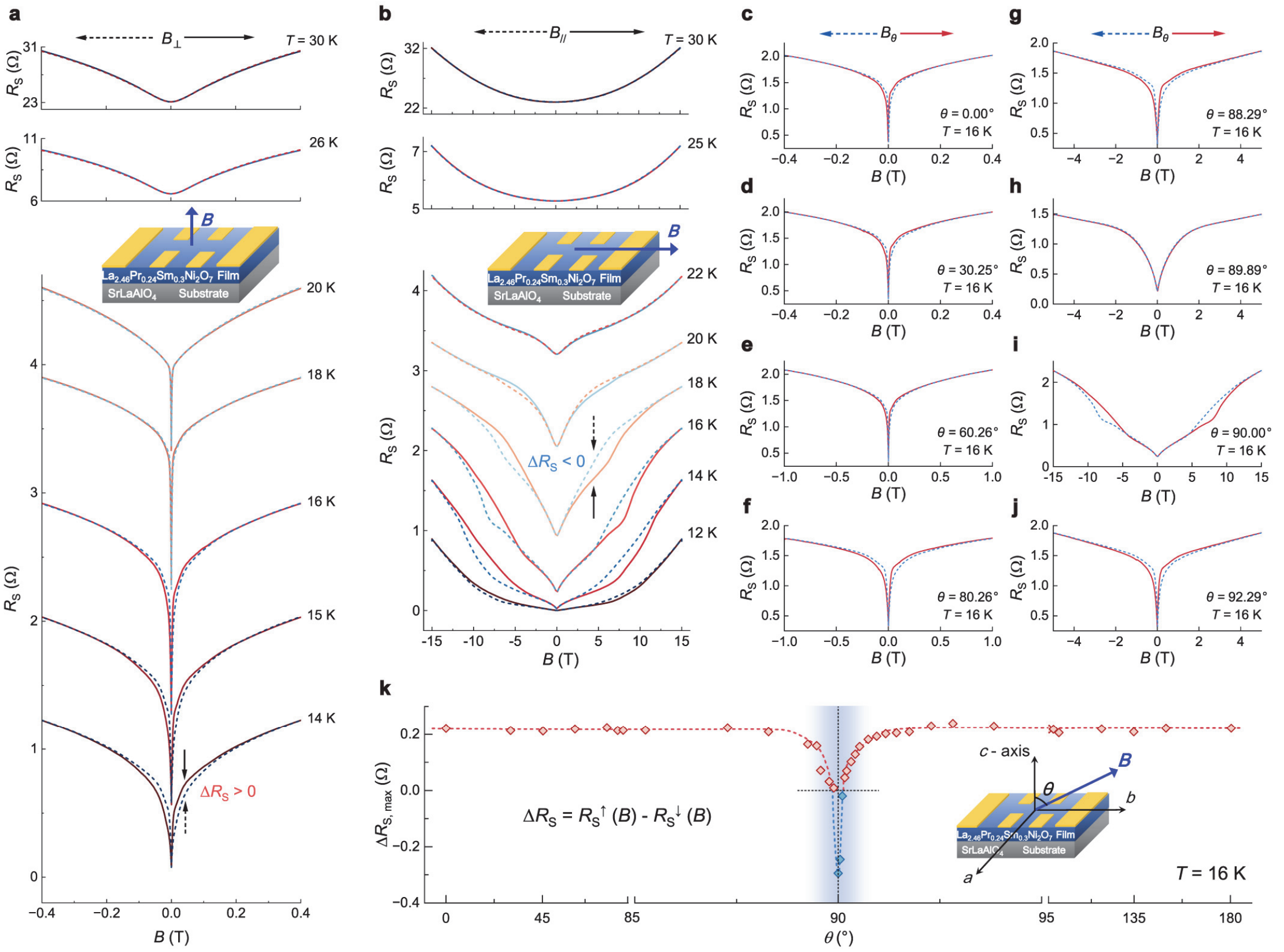

**Extended Data Fig. 3 | Hysteretic magnetoresistance in $La_{2.46}Pr_{0.24}Sm_{0.3}Ni_2O_7$ thin film (S4) under magnetic fields along different orientations. a**, $R_S(B)$ curves at different temperatures, zoomed-in to $|B_\perp| \leq 0.4$ T. Curves are offset vertically for clarity. Here, all $R_S(B)$ curves are measured with $B_\perp$ sweeping from 15 T to −15 T (dashed lines) and −15 T to 15 T (solid lines), while the selected range of $R_S(B)$ curves is displayed for clarity. The hysteresis is observed below 20 K and disappears at higher temperatures. **b**, $R_S(B)$ curves at different temperatures with $B_{/\!/}$ sweeping from 15 T to −15 T (dashed lines) and −15 T to 15 T (solid lines). The hysteresis is observed below 22 K and disappears at higher temperatures. **c-j,** Representative $R_S(B)$ curves at $T = 16$ K with the opposite swept magnetic fields $B_\theta$ along different polar angles $\theta$ from 0º (out-of-plane) to 90º (in-plane), and then to higher degrees, as denoted in the figures. **k,** $\theta$-dependence of $\Delta R_{S,\ max}$. Red and blue colors highlight the positive and negative polarities. Blue area and dashed lines are guides to the eye. The experimental resolution of $\theta$ is around 0.05º.

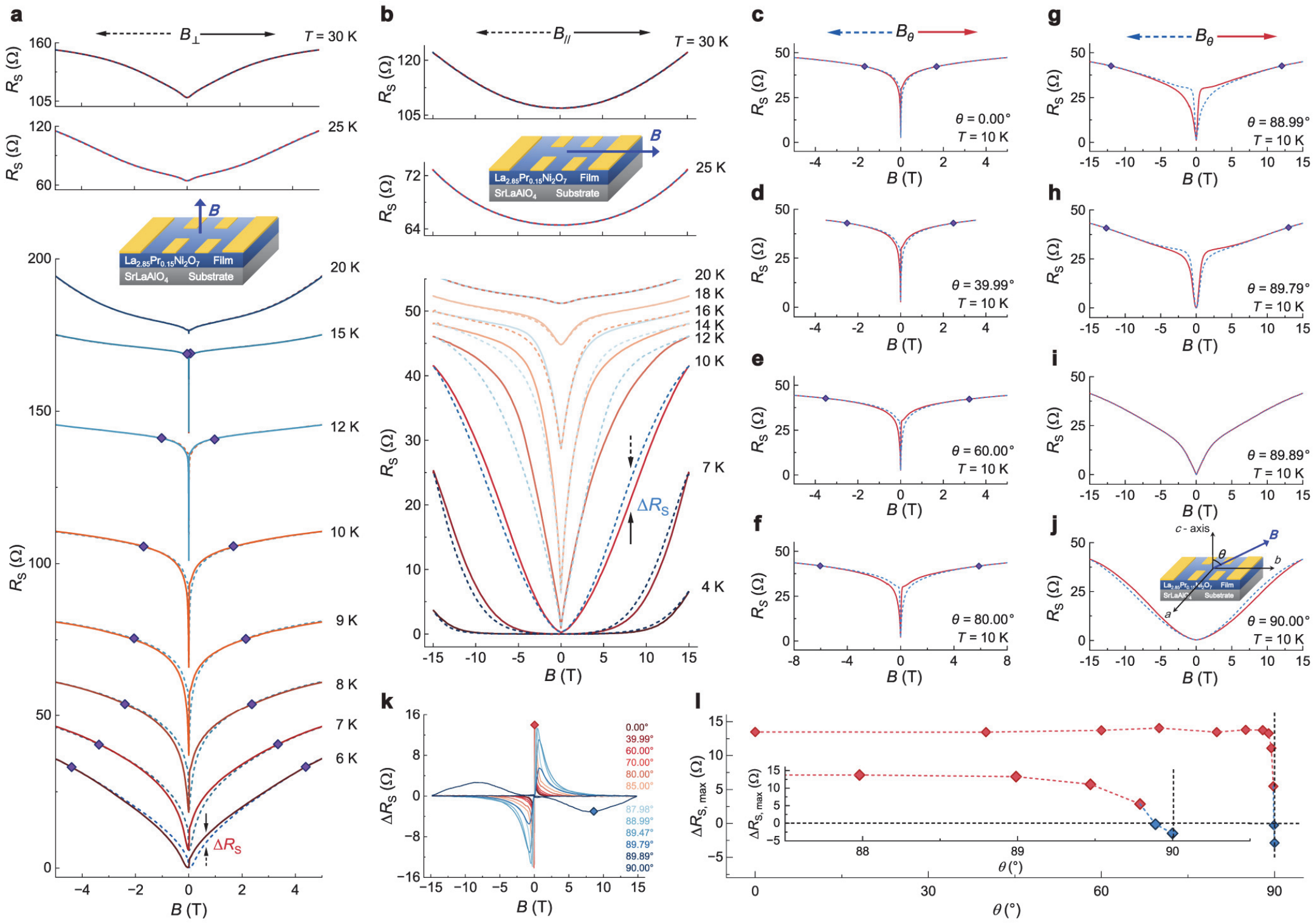

**Extended Data Fig. 4 | Hysteretic magnetoresistance in $La_{2.85}Pr_{0.15}Ni_2O_7$ thin film (S5) under magnetic fields along different orientations. a**, $R_S(B)$ curves at different temperatures, zoomed-in to $|B_\perp| \leq 5$ T. Curves are offset vertically for clarity. Here, all $R_S(B)$ curves are measured with $B_\perp$ sweeping from 15 T to −15 T (dashed lines) and −15 T to 15 T (solid lines), while the selected range of $R_S(B)$ curves is displayed for clarity. The hysteresis is observed below 15 K and disappears at higher temperatures. Purple diamonds indicate the saturation fields $B_{sat}$. **b**, $R_S(B)$ curves at different temperatures with $B_{//}$ sweeping from 15 T to −15 T (dashed lines) and −15 T to 15 T (solid lines). The hysteresis is observed below 18 K and disappears at higher temperatures. **c-j,** Representative $R_S(B)$ curves at $T$ = 10 K with the opposite swept magnetic fields $B_\theta$ along different polar angles $\theta$ from 0º (out-of-plane) to 90º (in-plane), as denoted in the figures. Purple diamonds denote the saturation fields. **k,** Anti-symmetrized $\Delta R_S(B)$ curves at different $\theta$ angles, where the $\Delta R_{S,\,max}$ at $\theta$ = 0º (defined as positive polarity, red diamond) and 90º (defined as negative polarity, blue diamond) are marked. **l,** $\theta$-dependence of $\Delta R_{S,\,max}$. Red and blue colors highlight the positive and negative polarities. Inset shows the zoomed-in view around $\theta$ = 90º (in-plane). Black dashed lines are guides to the eye.

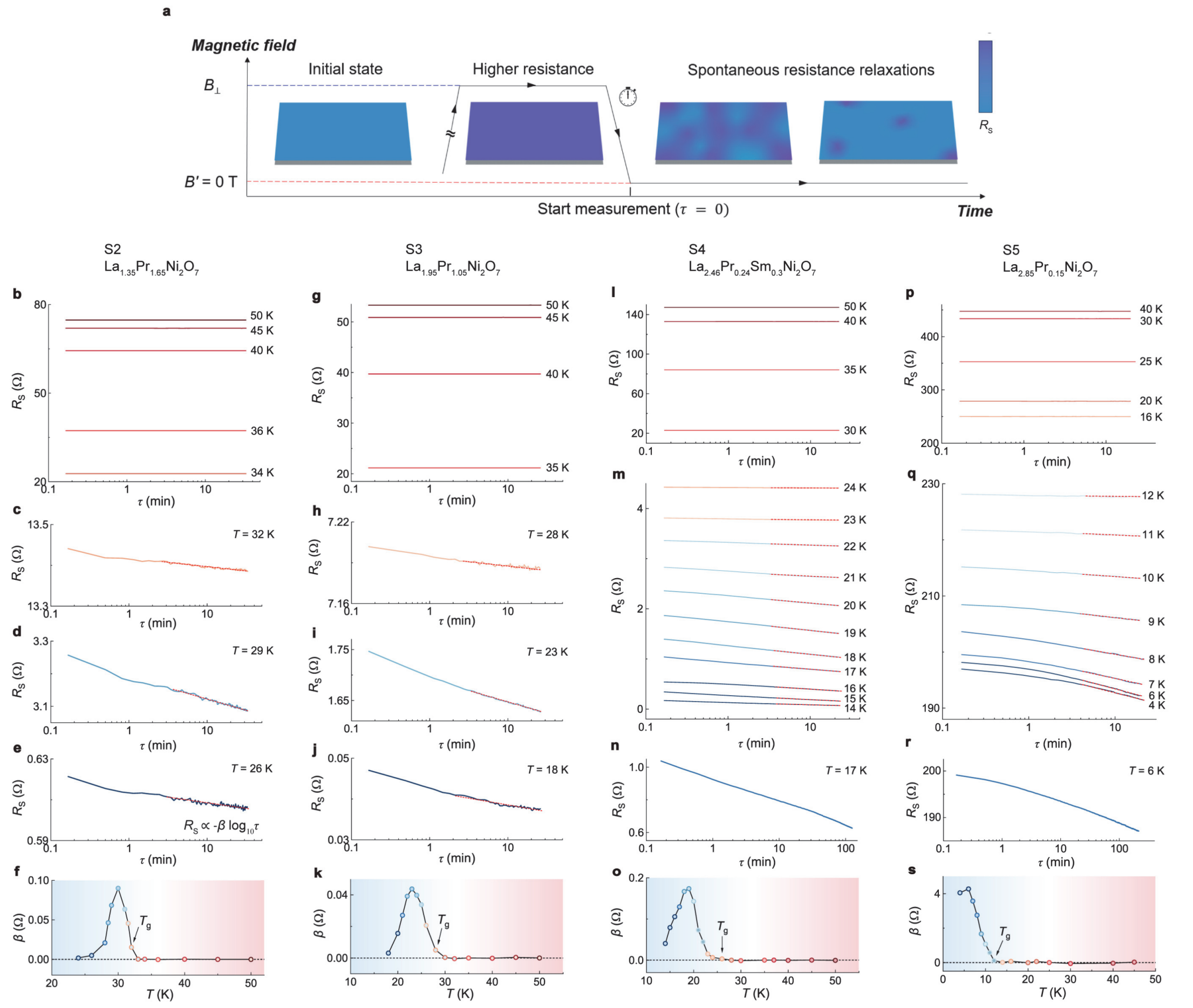

**Extended Data Fig. 5 | Time-dependent resistance relaxations in different (La, Pr, Sm)$_3$Ni$_2$O$_7$ thin films. a**, Schematic illustration of the measurement process. Columns from left to right show the time-dependent sheet resistance $R_S(\tau)$ at different temperatures and the extracted coefficient $\beta$ in La$_{1.35}$Pr$_{1.65}$Ni$_2$O$_7$ thin film (S2, **b** to **f**), La$_{1.95}$Pr$_{1.05}$Ni$_2$O$_7$ thin film (S3, **g** to **k**), La$_{2.46}$Pr$_{0.24}$Sm$_{0.3}$Ni$_2$O$_7$ thin film (S4, middle, **l** to **o**), La$_{2.85}$Pr$_{0.15}$Ni$_2$O$_7$ thin film with $p$ = 0.48 (S5, right, **p** to **s**), respectively. The magnetic field is applied and removed rapidly along the out-of-plane direction.

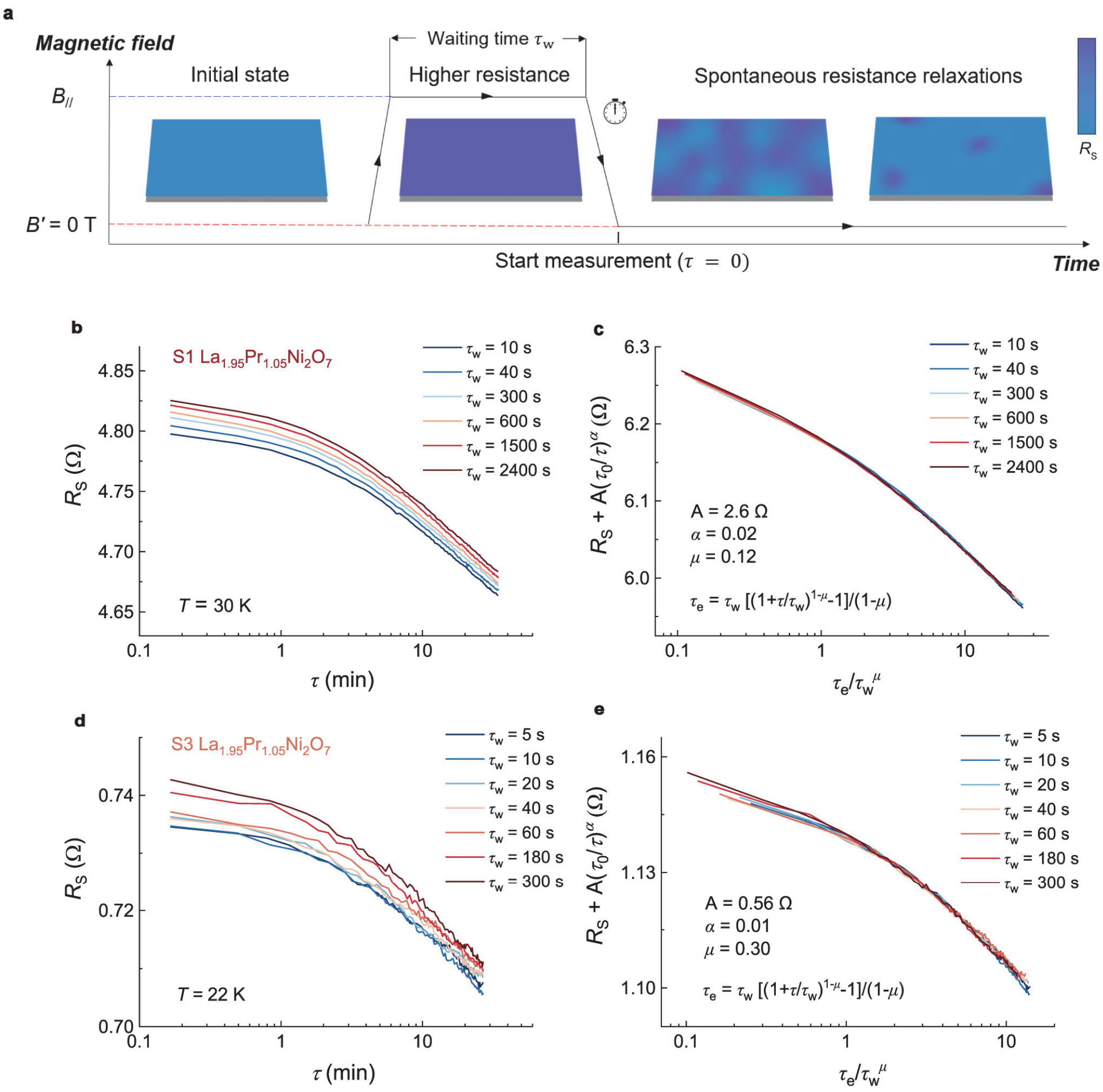

**Extended Data Fig. 6 | $\tau_w$-scaling analysis of $R_S(\tau)$. a,** Schematic illustration of the measurement process. Here, $B_{//}$ of 5 T is applied, held for a waiting time ($\tau_w$), and then removed rapidly. **b, c,** Time-dependent sheet resistance $R_S(\tau)$ at 30 K under different $\tau_w$ (**b**) and the scaling results (**c**) in $La_{1.95}Pr_{1.05}Ni_2O_7$ film (S1). **d, e,** $R_S(\tau)$ at 22 K (**d**) and the scaling results (**e**) in $La_{1.95}Pr_{1.05}Ni_2O_7$ film (S3). The scaling analysis plots the aging part of the sheet resistance, $R_S + A(\tau_0 / \tau)^\alpha$, as a function of $\tau_e / \tau_w^\mu$, where A is a constant, $\tau_0 \simeq 10^{-12}$ s is the microscopic attempt time, $\alpha$ and $\mu$ are scaling parameters, $\tau_e$ is effective time determined by $\tau_e = \tau_w[(1 + \tau / \tau_w)^{1-\mu} - 1] / (1 - \mu)$[72,73]. The parameters are denoted in the figure.

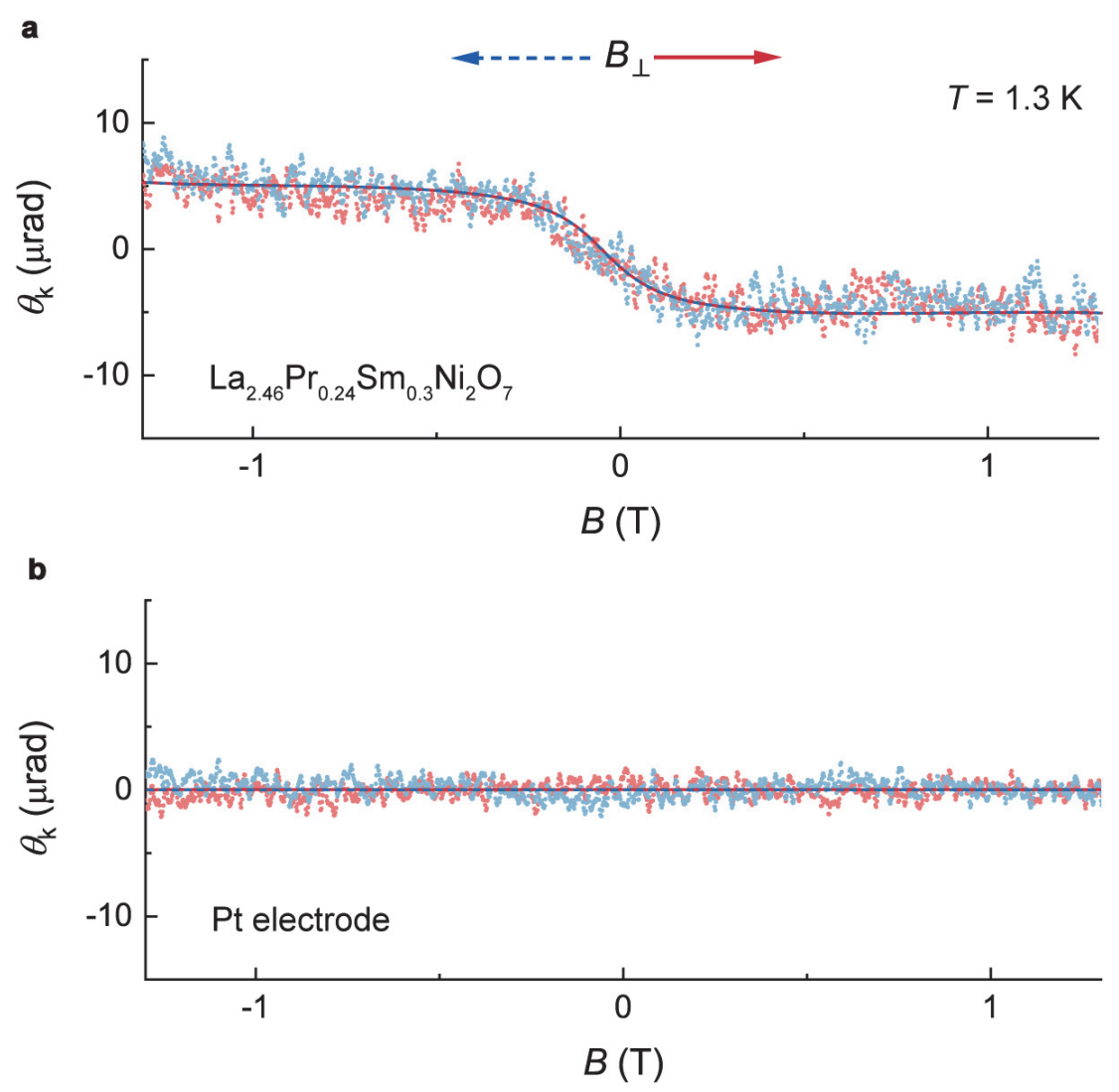

**Extended Data Fig. 7 | Magneto-optical measurements on $La_{2.46}Pr_{0.24}Sm_{0.3}Ni_2O_7$ film (S4).** Magneto-optical Kerr angles ($\theta_K$) measured on the $La_{2.46}Pr_{0.24}Sm_{0.3}Ni_2O_7$ film (**a**) and the Pt electrode on the film (**b**) at 1.3 K. The blue dashed lines and red solid lines are guides to the eyes. The magneto-optic Kerr effect (MOKE) is measured using an all-fiber Sagnac interferometer at optical wavelength $\lambda$ = 1550 nm[74]. The finite MOKE signal requires TRS breaking, where the extracted Kerr angle $\theta_K$ is proportional to both the Hall conductivity and the magnetization[75]

$$\theta_K \propto \sigma_{xy}(\omega) \propto \pm M_z$$

The non-zero $\theta_K$ values and the magnetic field-dependent anomalies around zero field are observed in the $La_{2.46}Pr_{0.24}Sm_{0.3}Ni_2O_7$ film, providing the complementary evidence for TRS breaking. In contrast, no discernible field-dependence of $\theta_K$ is detected on the Pt electrode.

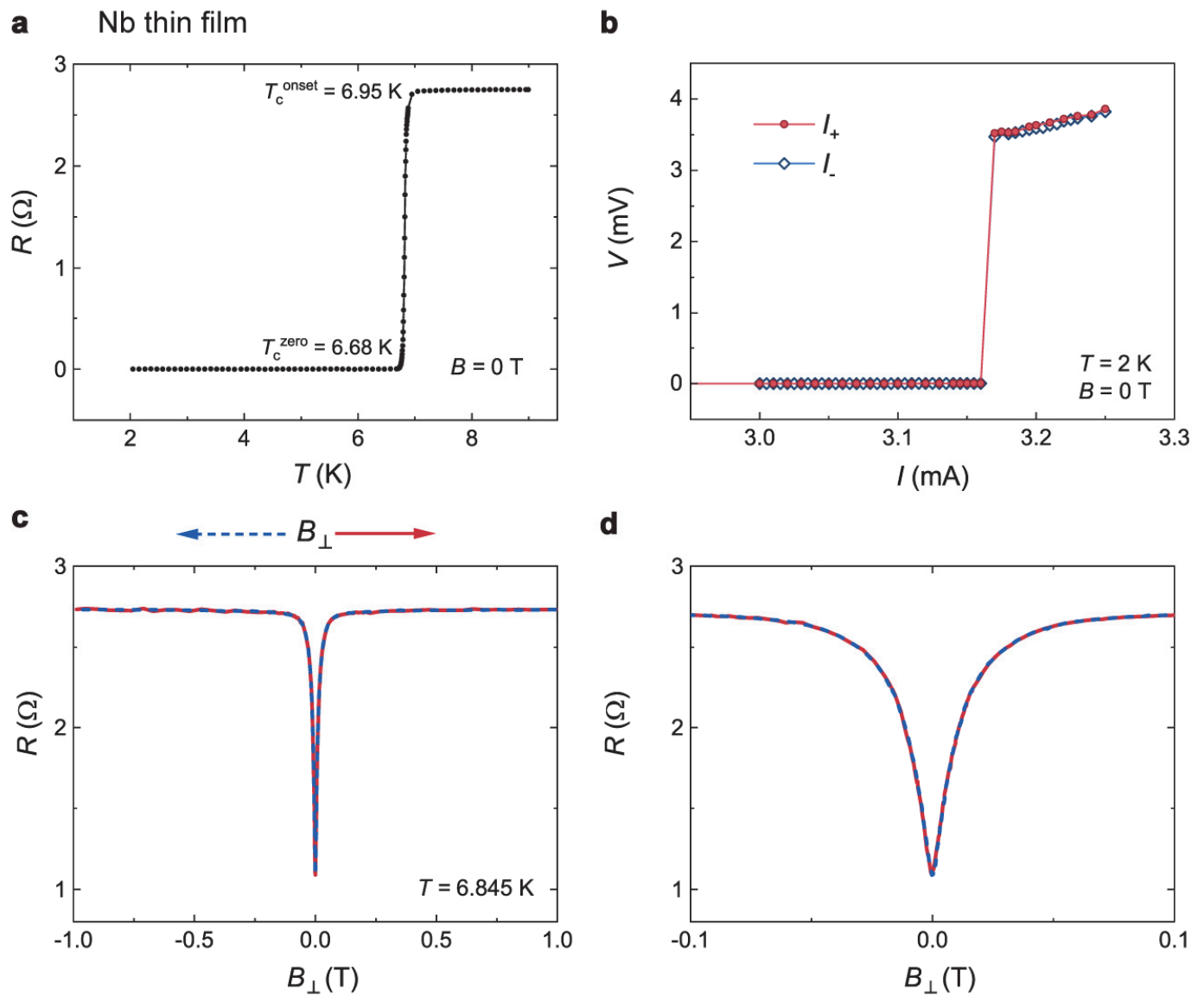

**Extended Data Fig. 8 | Absence of hysteresis in a 30-nm thick Nb thin film. a,** $R(T)$ curve of a 30-nm thick Nb thin film. **b**, $I$–$V$ curves at 2 K and 0 T, containing both $V(I_+)$ (red solid dot, current applied from zero to positive), $|V(I_-)|$ (blue hollow diamond, current applied from zero to negative). No discernible deviation between $V(I_+)$ and $|V(I_-)|$ curves is observed, indicating the absence of non-reciprocity. **c** and **d,** Magnetoresistance $R(B)$ curves under opposite swept magnetic fields at 6.845 K with different zoomed-in scales, where no hysteresis loop is observed. The Nb thin film and the electrodes with Hall bar geometry are fabricated using the standard photolithography technique. The Nb thin film is grown by magnetron sputtering. The electrodes are grown by electron beam evaporation.